\documentclass{article}

\usepackage{arxiv}

\usepackage[utf8]{inputenc} 
\usepackage[T1]{fontenc}    
\usepackage{hyperref}       
\usepackage{url}            
\usepackage{booktabs}       
\usepackage{amsfonts}       
\usepackage{nicefrac}       
\usepackage{microtype}      
\usepackage{lipsum}
\usepackage{graphicx,verbatim,float,subfigure,graphicx,color,siunitx}
\usepackage{amsmath}
\graphicspath{ {./images/} }

\usepackage{amssymb}
\usepackage{svg}
\usepackage{multirow}  
\usepackage[export]{adjustbox}
\usepackage{soul}
\usepackage{textcomp}
\usepackage{multirow}
\usepackage{enumitem}
\usepackage[sort&compress,numbers]{natbib}

\title{Fiber Packing and Morphology Driven Moisture Diffusion Mechanics in Reinforced Composites}

\author{
 S.P. Subramaniyan, M.A. Imam and  P. Prabhakar$^*$ \\
  Dept. of Civil \& Environmental Engineering \\
  University of Wisconsin-Madison \\
  Madison, WI 53706  \vspace{0.05in} \\
  \texttt{$^*$pavana.prabhakar@wisc.edu}
}

\begin{document}
\maketitle
 
\newcommand{\SPS}[1]{\textcolor{red}{\bf{Sabari: #1}}}

\begin{abstract}
Fiber reinforced polymer composite (FRPC) materials are extensively used in lightweight applications due to their high specific strength and other favorable properties including enhanced endurance and corrosion resistance. However, these materials are inevitably exposed to moisture, which is known to drastically reduce their mechanical properties caused by moisture absorption and often accompanied with plasticization, weight gain, hygrothermal swelling, and de-bonding between fiber and matrix. Hence, it is important to understand the mechanics of moisture diffusion into FRPCs. The presence of fibers, especially impermeable like Carbon fibers, introduce tortuous moisture diffusion pathways through polymer matrix. In this paper, we elucidate the impact of fiber packing and morphology on moisture diffusion in FRPC materials. Computational models are developed within a finite element framework to evaluate moisture kinetics in impermeable FRPCs. We introduce a tortuosity factor for calculating the extent of deviation in moisture diffusion pathways due to impermeable fiber reinforcements. Two-dimensional micromechanical models are analyzed with varying fiber volume fractions, spatial distributions and morphology to elucidate the influence of internal microscale fiber architectures on tortuous diffusion pathways and corresponding diffusivities. Finally, a relationship between tortuosity and diffusivity is established such that diffusivity can be calculated using tortuosity for a given micro-architecture. Tortuosity can be easily calculated for a given architecture by solving steady state diffusion governing equations, whereas time-dependent transient diffusion equations need to be solved for calculating moisture diffusivity. Hence, tortuosity, instead of diffusivity, can be used in future composites designs, multi-scale analyses, and optimization for enabling robust structures in extreme moisture environments.
\end{abstract}







\section{Introduction}\label{intro}
Fiber reinforced polymer composite (FRPC) materials are widely used in structural applications within the aerospace, automotive, civil and marine industries due to their high specific strength, improved durability and corrosion resistance. When used in such applications, FRPCs are inevitably exposed to extreme conditions: moisture, low and high temperatures, pH, UV radiation and salinity, among others. Moisture, in particular, is a key environmental factor due to 1) its prevalence, although to different extents; 2) polymer matrices having a relatively high polarity when compared to reinforcing fibers and therefore a high affinity for absorbing moisture in humid environments; 3) the drastic reduction of FRPC properties caused by moisture absorption and often accompanied with plasticization, weight gain, hygrothermal swelling, and de-bonding between fiber and matrix \cite{Dattaguru1986,Whitcomb2002}. This necessitates an in-depth study of the mechanisms and avenues of moisture ingression through FRPCs. 


Most prior experimental research conducted on FRPCs for establishing the effect of moisture diffusion involves measuring the weight gain response with various reinforcing fiber types (continuous or textile) and materials (Carbon, Glass, Kevlar) \cite{Bao2002,Pavlidou2005,Marcovich1999,Barjasteh2012,Cavasin2019,Taghavi2000,Perez-Pacheco2013,Ray2006,Dan-Mallam2015,Gao2019,Zenasni2006,Guloglu2020,Surathi2006,Dana2013,Airale2016,ElSawi2014}. In most studies, FRPC samples are subjected to moisture (chamber, submerged in water/seawater) and their weight gain is measured periodically, and further plotted as weight gain vs. time graphs. Key information obtained are the effective diffusivity and moisture saturation for specific FRPCs. Such studies are very important for identifying the overall effect of moisture diffusion and the extent of moisture absorption, although they only provide a global understanding of moisture diffusion and the local mechanisms are poorly understood or rarely captured. To understand these, other researchers have developed computational models within the finite element method to simulate moisture diffusion through FRPCs \cite{Pan2019,Wong2016,DeBrito2019,Gagani2018,Korkees2018,Kostopoulos2017,Pasupuleti2011,Laurenzi2008,Tang2005,Bond2005,Fichera2015,Yang2010,Huo2016,Roe2012,Gueribiz2013,Chu2019,Zheng2019}. These frameworks have focused on predicting experimentally measured moisture diffusivities and global weight gain curves of FRPCs considering independent parameters like fiber volume fraction, fiber distribution and diffusivities of individual components (fiber and matrix). However, gaining a fundamental understanding of micro-scale moisture diffusion in terms of the effective fiber architecture has seldom been the focus. 

Bond et al.~\cite{Bond2005} investigated the influence of fiber spatial distribution on moisture diffusivity computationally and stated that diffusion coefficient is unaffected between regular and random fiber distribution. Hence, they suggested that the regular fiber model is adequate for estimating diffusion coefficients for composites. However, Joliff et al.~\cite{Joliff2012a} showed that fiber arrangement has a significant impact on moisture diffusion when the fibers are in contact with each other, thereby, forming an effective barrier against diffusion. The results also showed that the rate of diffusion is greatly modified during initial (short) exposure times because of fiber arrangement, where the diffusivity of randomly distributed array is lower compared to that of a regular array of fibers.
Aditya et al.~\cite{Aditya1993} showed that moisture diffusivity is influenced by the cross-sectional shape of fibers. They computationally showed that the effective diffusivity is minimum for circular fibers, and it increases as the eccentricity of circular cross-section increases to become an elliptical cross-section. In their consecutive work~\cite{Aditya1994}, they considered computational models with many irregular shapes of fiber cross-section and determined that the  percentage change in diffusivity when compared with circular fibers varies between -15.19\text{\%} to 10.56\text{\%}. In summary, prior research has shown that moisture diffusivity of fiber reinforced composites depends on fiber packing, fiber volume fraction, fiber cross-sectional shape and diffusion properties. Although these parameters are studied individually, further investigation is necessary to study the combined effect of these microscale fiber architecture on moisture diffusivity.

In this paper, we propose to capture the combined effect of fiber micro-architecture by introducing a tortuosity factor. Such a factor has previously been considered for capturing diffusive flow in porous electrodes ~\cite{Shen2007}. In this paper, we will numerically determine the effective diffusivity and compare that with the tortuosity factor to account for the influence of fiber architecture on moisture diffusion in FRPCs. Particularly, we used Finite Element Method (FEM) for studying the moisture ingression in impermeable fiber reinforced polymer composites. First, we developed two-dimensional micromechanical models with different fiber volume fractions and fiber distributions to establish the influence of fiber distribution and volume fraction on the effective moisture diffusivity and to determine the tortuosity factor for these composites. Then, we studied the influence of fiber morphology (shapes) on tortuosity. Finally, we established a relationship between tortuosity that captures the fiber architecture (combined effect of fiber volume fraction, fiber distribution and fiber shape) and diffusivity. This relationship was validated for a set of test fiber micro-architectures.

\section{Motivation and Significance} \label{se:motivation}
The main objective of this paper is to study how moisture diffusion occurs in FRPC composites. In particular, our model system is Carbon fiber reinforced polymer matrix (CFRP) composites, where Carbon fibers are considered to be impermeable to moisture diffusion. Hence, they typically act as barriers that make moisture diffusion pathways tortuous, thereby, delaying moisture ingression and consequently reducing moisture diffusivity. Fiber spatial distribution and morphology induced hindrance to moisture diffusion is quantified using a tortuosity factor, where an increase in tortuosity implies greater resistance to moisture diffusion. The tortuosity is obtained by solving steady state diffusion governing equation, which is computationally less expensive compared to transient diffusion equation that is solved for obtaining moisture diffusivity. Hence, by establishing a relationship between tortuosity and diffusivity, we can directly obtain diffusivity from tortuosity values for a large range of fiber architecture without having to perform diffusion analyses. This is particularly attractive within multi-scale analysis and optimization problems. 

\begin{figure}[h!]
 \centering
  \includegraphics[width=12cm]{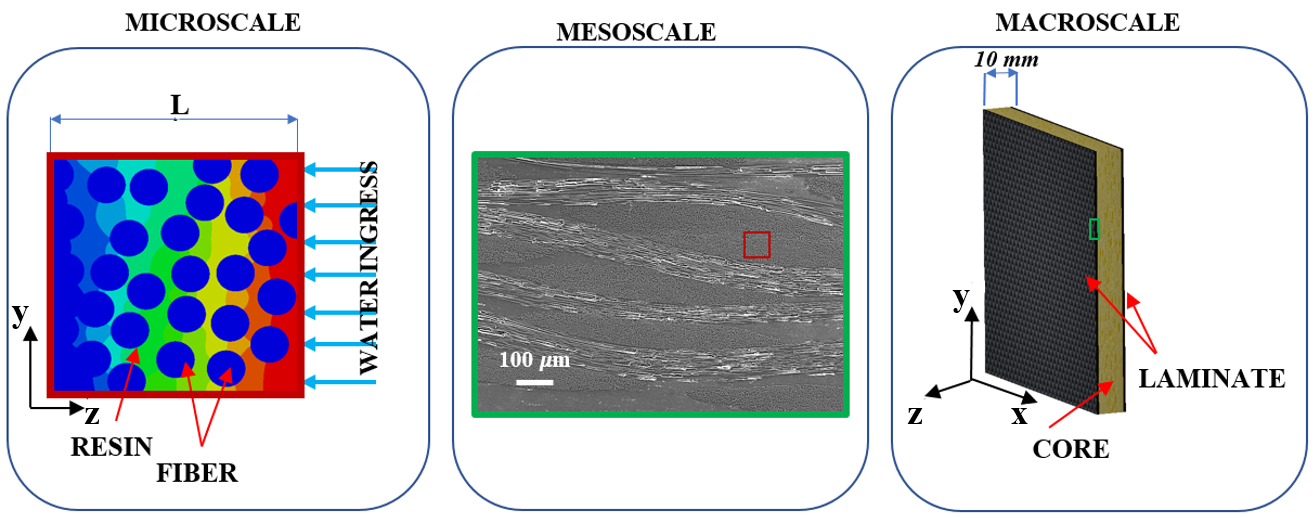}
  \caption{Macroscale image (right) consists of a sandwich structure composed of Carbon fiber reinforced plain weave polymer composite facings with foam core. Mesoscale image (center) shows a Scanning Electron Microscopy image of the cross-section of one of the facings. Microscale image (left) shows the representative volume element model of a fiber bundle present within a facing. Exposure to moisture occurs on either of the facings along to the thickness direction. As a result, at the microscale, moisture diffusion occurs primarily in the direction perpendicular to the fiber direction. }\label{macro_micro}
\end{figure}

Further in this study, we consider one dimensional diffusion while modelling moisture kinetics. This is because FRPCs are typically used as large thin sheets or plates where thickness is very negligible compared to the in-plane dimensions. When used in structures, the faces are exposed to moisture which results in primary moisture flow direction occurring along the through-thickness direction, as explained in Figure~\ref{macro_micro}.

In summary, the novelty of this paper is twofold:

\begin{itemize}
    \item We present a novel tortuosity factor within the realm of FRPCs that can capture the collective influence of fiber architecture. Although presented for 2D micromechanics models in this paper, the concept can be extended to 3D space and multiscale architectures. For example, our work can be extended to multi-scale analysis of 2D and 3D textile composites where the effect of weave patterns, tow cross-sections, tow waviness, etc. on moisture diffusion can be captured effectively by relating diffusivity and tortuosity.
    
    \item By establishing a relationship between tortuosity and diffusivity, diffusivity can be readily estimated using tortuosity. Further, we can use tortuosity as a design parameter for FRPCs composites, within multi-scale analysis, and optimization problems instead of diffusivity. Such tortuosity factors can also be developed for calculating effective electrical and thermal properties.
    
\end{itemize}

\section{Computational Models}\label{se:compModel}
Finite element method is used here for studying moisture diffusion in FRPCs. Two-dimensional  representative volume element (RVE) micromechanical models are developed for understanding the effect of fiber distribution and fiber volume fraction on moisture diffusivity and tortuosity. These models consist of different spatial distribution of fibers in polymer matrix with a range of fiber volume fractions. Then, the influence of fiber morphology/shape on moisture diffusion is elucidated using a unit cell model. Finally, the combined effect of fiber architecture on moisture diffusivity and tortuosity is studied by modelling RVEs of real microstructures of FRPCs. Although the interphase between fibers and matrix could impact moisture diffusion, we have only considered two-phase (fiber and matrix) models in this paper due to lack of sufficient characterization of interphase chemistry in the literature. Having said that, the approach and models can be easily extended to three-phase models to include fiber-matrix interphases.

In this paper, the influence of fiber architecture on moisture diffusion and tortuosity is investigated first in two steps: 1) fiber volume fraction and spatial distribution, and 2) fiber morphology. Next, real microstructures from FRPCs are considered to capture the combined effect of fiber architecture on moisture diffusion and tortuosity. 

\begin{figure}[H]
\centering
\subfigure[]{
\includegraphics[width=12cm]{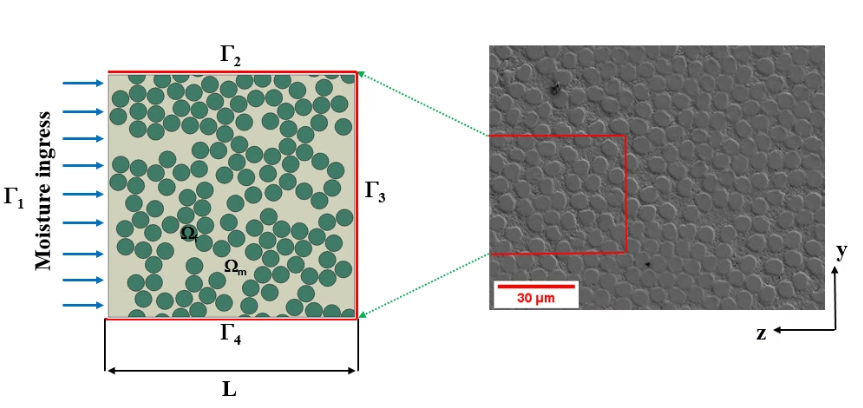} \label{schematic}
}
\hfill
\centering
\subfigure[]{
\includegraphics[width=12cm]{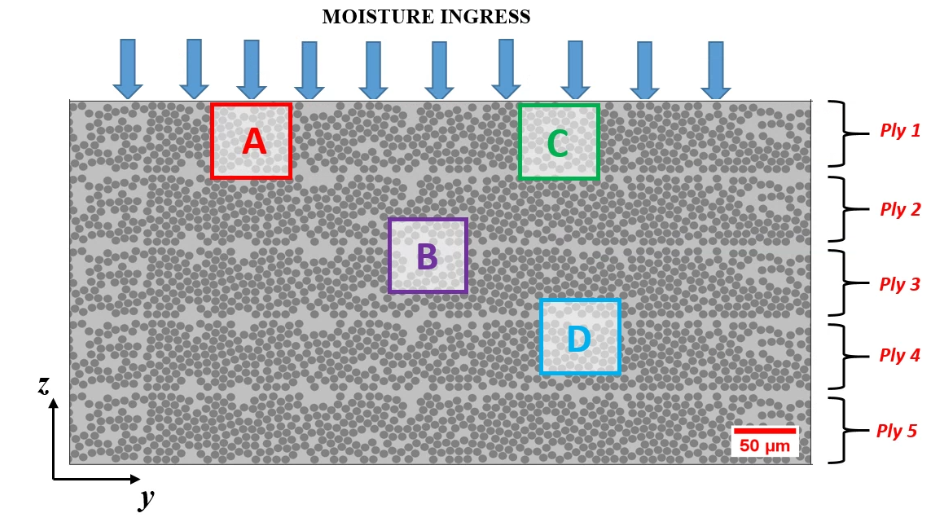}\label{Rve type}
}
\caption{(a) Illustration of a 2D micromechanical model (left) with material domains and boundaries highlighted. Cross-sectional view of a real microstructure (right) within a CFRP composite at the microscale, and (b) Representation of different types of RVEs with (A, C) and without (B, D) boundary effect.  }\label{fig:RVE_material_periodicity}
\end{figure}

{\bf Fiber volume fraction and spatial distribution:} 2D micromechanical finite element models of RVEs with spatially distributed fibers are considered as shown in Figure~\ref{schematic}.  Figure~\ref{Rve type}shows a representative cross-sectional image of a carbon fiber-reinforced polymer laminate with five plies where the dimensions along the y-axis are larger than dimensions along the z-axis. Here, A-D shows possible RVE types. In general, RVEs like A and C are not considered while modeling the effective elastic properties due to the boundary effect, and only RVEs like B and D which are in the bulk of the material are considered \mbox{\cite{Gitman2007}}. In our case, since the interest is in modeling the moisture diffusion from the boundary to the interior of the material, we consider RVEs like A and C that are at the boundary. In this work, only the moisture ingression that occurs along the thickness direction is considered since the composite's length and width are orders of magnitude larger than the thickness. Therefore, the material periodicity in our 2D models is considered only in the y-direction for numerical simulations. Although, geometric periodicity is considered along the y-axis, we have not considered periodic boundary conditions in our models. Typically, there are three types of boundary conditions used for numerical homogenization in diffusion (conduction) problems, which are uniform concentration gradient boundary condition, uniform flux boundary condition, and periodic boundary condition. In this work, we have considered uniform concentration gradient boundary condition to calculate the tortuosity, which is the average flux along the z-axis, while the other surfaces are insulated. This process needs to be repeated depending upon the direction of interest. Further, these three boundary conditions converge to the same effective properties when the model is sufficiently large \mbox{\cite{Yvonnet}}. In our model, the size of the RVE is large enough such that the influence of either of the three types of boundary conditions mentioned above on the effective properties are minimal. Next, we discuss how we determined the converged size for our RVE. 

\begin{figure}[H]
 \centering
  \includegraphics[width=8cm]{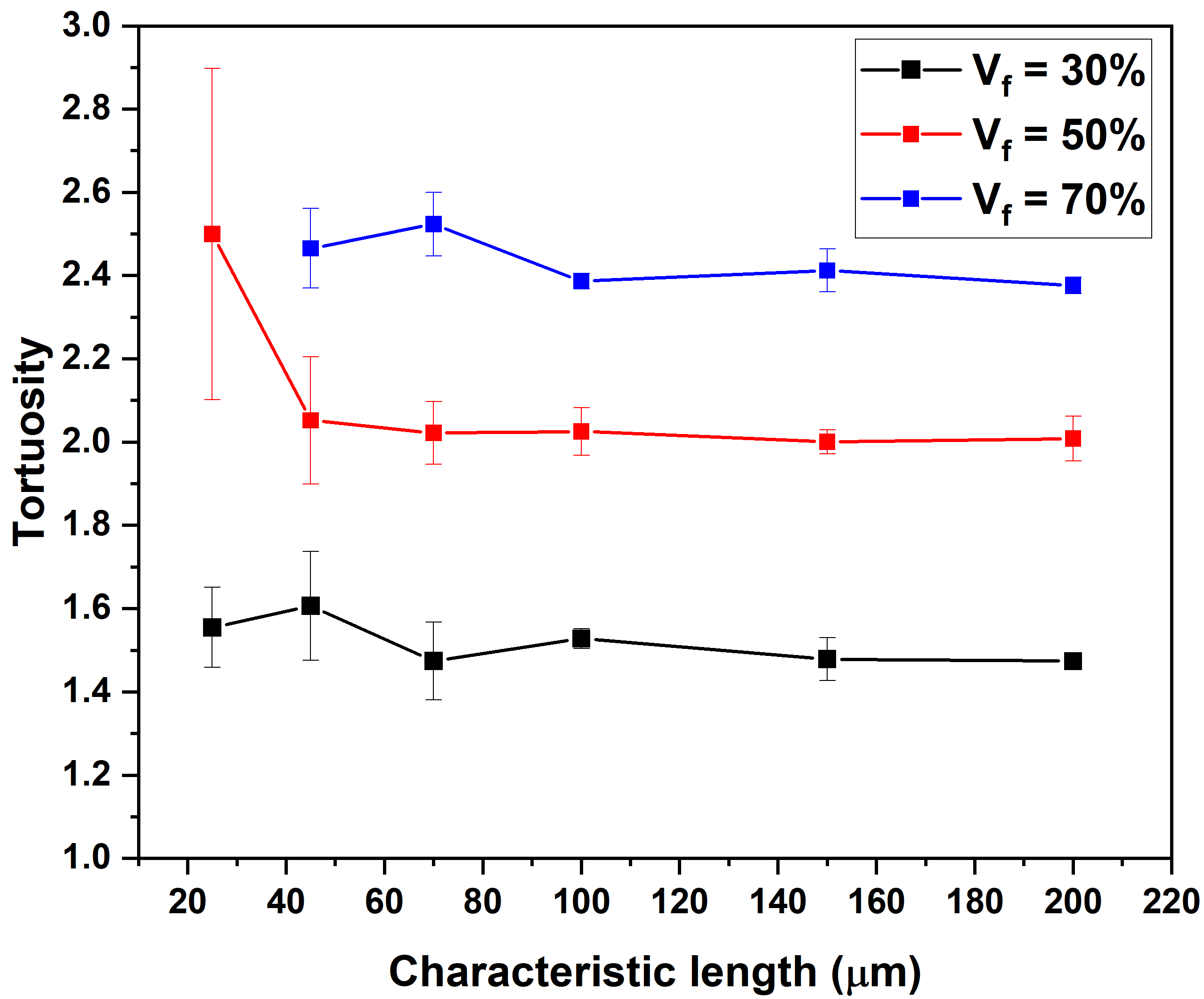}
  \caption{Convergence study to determine the RVE size with five realizations for each characteristic length and three different volume fractions.}\label{Rve_size_determination}
\end{figure}

The overall domain of the models, $\Omega$, is the union of fibers and matrix regions given by $\Omega_f+\Omega_m$. The dimensions of the entire domain, $\Omega$, is 100$\mu m$ x 100 $\mu m$ and the fiber diameter is 7$\mu m$. The exterior boundaries of the domain are $\Gamma_i$, where $i=1-4$. We used a numerical statistical approach to determine the size of the RVE, where 2D square models with edge (characteristic) length of 25, 45, 70, 100, and 200 $\mu m$ are first considered. For every RVE size mentioned above, five random fiber distributions each are generated for three fixed fiber volume fractions (30, 50 and 70 \%). We then determined the effective diffusion properties in terms of tortuosity as described later in \mbox{section \ref{tortuosity_calculation}}. The mean value and standard deviation for all the RVEs for each volume fractions considered are calculated and plotted in \mbox{Figure~\ref{Rve_size_determination}}. We notice that the mean value of tortuosity converges to a plateau as the sample size increases. From this approach, RVE sample size of 100 $\mu m$ x 100 $\mu m$ is considered, which has less than 3\% error compared to the largest RVE size.

\begin{figure}[h!]\label{fiber spatial}
\centering
\subfigure[]{
\includegraphics[width=4cm]{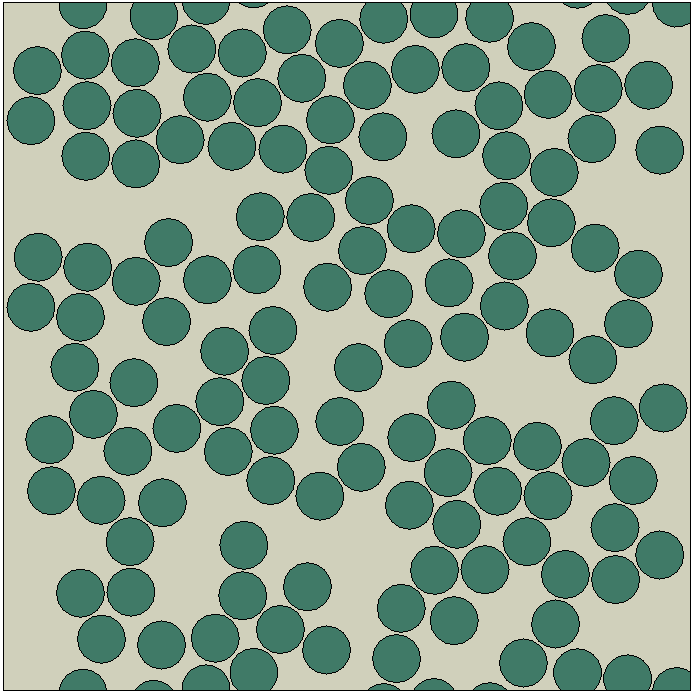} \label{random_array}
}
\hfill
\centering
\subfigure[]{
\includegraphics[width=4cm]{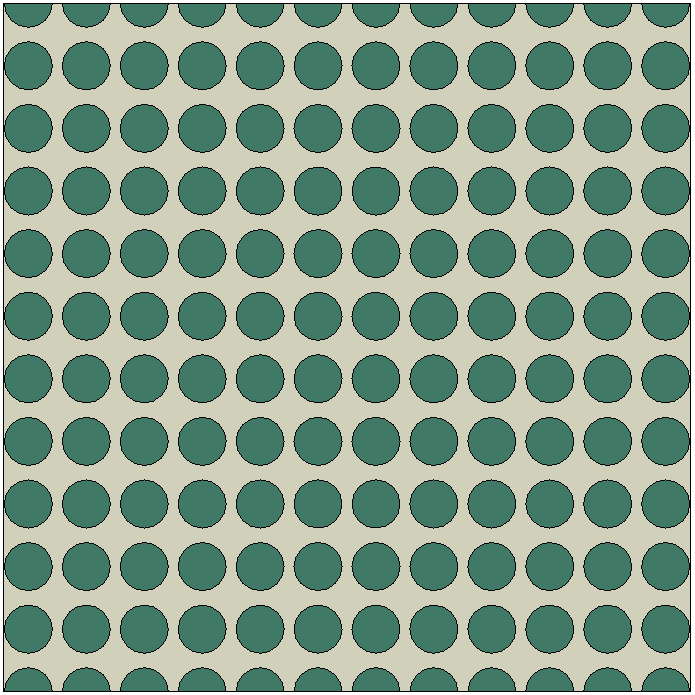}\label{square_array}
}
\hfill
\centering
\subfigure[]{
\includegraphics[width=4cm]{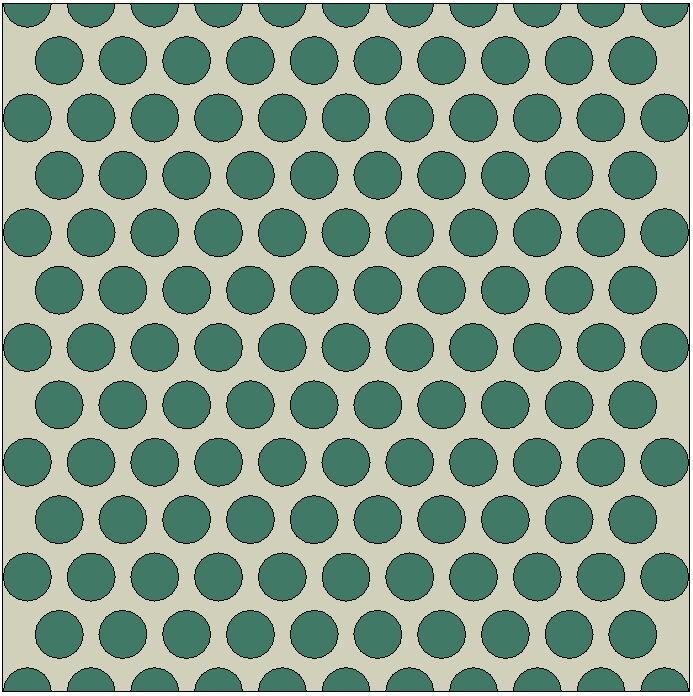}\label{hex_array}
}
\caption{Spatial distribution of fibers (green) in matrix (cream): (a) Random array; (b) Square array; (c) Hexagonal array}
\end{figure}

Extensive prior research has been performed using regular arrays of fiber distribution or by considering a unit cell in a 2D micromechanical model~\cite{Whitcomb2002}~\cite{Gueribiz2009} to determine the effective diffusivity of composites. In this paper, our model comprises of random fiber distribution and different volume fractions that closely captures the essence of the microstructure of a real composite shown in Figure~\ref{schematic} (right). One of the primary reasons for choosing such a representation is to emphasize the vital role of volume fraction, fiber distribution and associated tortuosity on moisture ingression in composite materials reinforced with fibers. {\em We hypothesize that the tortuosity is enhanced by higher fiber volume fractions and more disorderliness in fiber arrangement, which reduces the effective diffusivity of moisture ingression into composites, thus, increasing the time needed for complete degradation of the material.} Therefore, 2D models with different fiber distributions and volume fractions are created as shown in Figures~\ref{random_array}, ~\ref{square_array} and ~\ref{hex_array}. Randomly distributed models with high fiber volume fractions like 60, 65 and 70\text{\%} are created using a microstructure generation tool - VIPER \cite{Herraez2020}. These models are then imported into finite element framework for further analysis. A transient moisture diffusion boundary value problem is solved within the finite element framework in ABAQUS~\cite{abaqus} to determine the effective diffusivity of each composite model considered. This time-dependent moisture diffusion is based on Fick’s second law, which will be discussed later in Section~\ref{diffusionModeling}.

\begin{figure}[h!]
\centering
\subfigure[]{
\includegraphics[width=3cm]{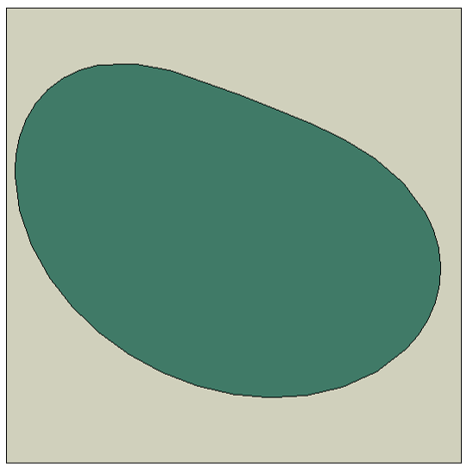} \label{a_1_b_3}
}
\hfill
\centering
\subfigure[]{
\includegraphics[width=3cm]{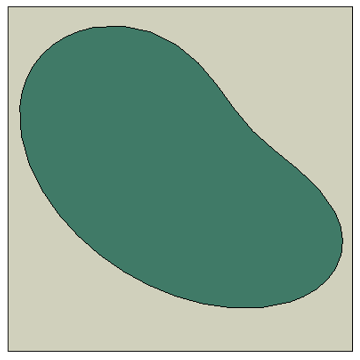}\label{a_1_b_5}
}
\hfill
\centering
\subfigure[]{
\includegraphics[width=3cm]{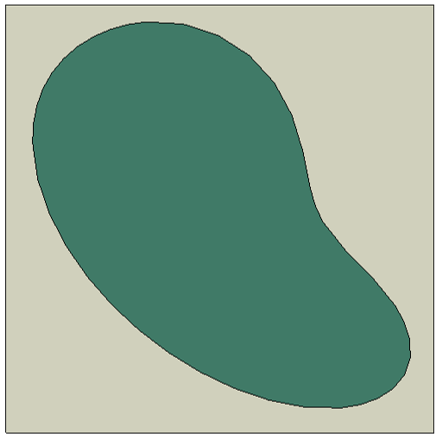}\label{a_1_b_7}
}
\hfill
\centering
\subfigure[]{
\includegraphics[width=3cm]{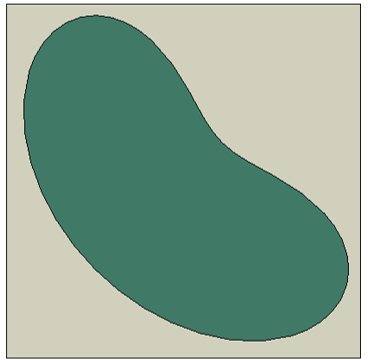}\label{a_3_b_3}
}
\caption{Unit cell models with constant fiber cross-sectional area, but with different fiber morphology parameters $a$ and $b$: (a) a=1, b=3 ; (b) a=1, b=5; (c) a=1, b=7; (d) a =3, b=3}\label{morphologyEquation}
\end{figure}

{\bf Fiber morphology:} The morphology of fiber cross-section in FRPCs can play an important role on moisture diffusivity and tortuosity\cite{Aditya1993,Aditya1994}. In order to gain an in-depth understanding of the influence of fiber morphology on tortuosity, the morphological features have been further subdivided into 1) fiber cross-sectional perimeter (2D) or fiber cross-sectional surface area (3D) and 2) fiber orientation. For the fiber perimeter study, different fiber shapes like square, asteroid, octagon and circle are chosen such that the tortuosity is the same along the x and y directions. Then, to study the influence of fiber cross-section angle, an elliptical cross-section is chosen and the tortuosity is determined for different fiber cross-sectional angle starting at 0 degrees to 90 degrees of the major axis to the x axis. However, since the real microstructure is a combination of circular and deformed shapes, different bean curved cross-sectional shapes are considered using the following parametric equations: 
\begin{equation}\label{beanCurveEquation}
\begin{split}
    x= r*cos(t)(sin^a(t)+cos^b(t))\\
   y= r*sin(t)(sin^a(t)+cos^b(t)) 
   \end{split}
\end{equation}
where, r is the radius, a and b are morphology parameters. Few unit cell models with deformed fiber cross-sections having constant cross-sectional area and different morphology parameters are shown in Figure~\ref{morphologyEquation}.
 
\begin{figure}[H]\label{real_microstructure_modelling}
\centering
\subfigure[]{
\includegraphics[width=4.7cm]{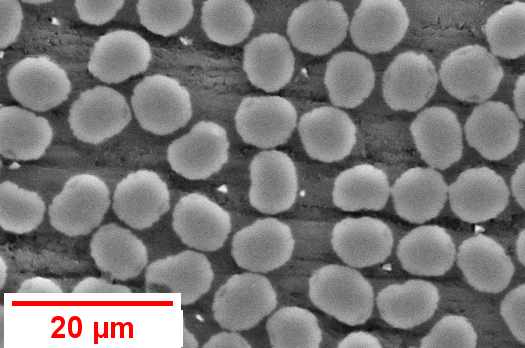} \label{real_micro_SEM}
}
\centering
\subfigure[]{
\includegraphics[width=4.8cm]{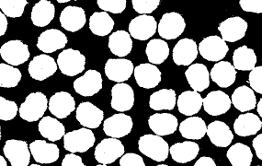}\label{real_micro_two_phase}
}
\centering
\subfigure[]{
\includegraphics[width=4.8cm]{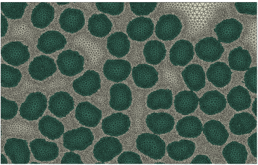}\label{real_micro_mesh}
}
\caption{Process of converting real microstructure image of FRPC into a finite element meshed model: (a) SEM image of a CFRP composite; (b) Two phase binary image; (c) Meshed Finite Element Method model}
\end{figure}

{\bf Real microstructure:} 
Diffusivity and tortuosity of real microstructures within FRPCs are determined next by developing their microscale models. To that end, images of CFRP composites at the microscale are obtained using a Zeiss LEO 1550 VP (Carl Zeiss Jena GmbH–Planetariums, Germany) field emission scanning electron microscope (SEM) at UW-Madison. Specimen cross-sections are polished thoroughly using SiC grinding paper up to 1200 grit size followed by polishing using DACRON II polishing cloth with 1-0.1 mm diamond solution. Then, the cross-section of these specimens are examined under 15 kV electron beam and secondary electron detector to obtain different magnified images with high resolution. The polished gray scale cross-section images similar to the one shown in Figure~\ref{real_micro_SEM} from SEM are converted into two-phase binary images as shown in Figure~\ref{real_micro_two_phase} using an image processing tool. These binary images are imported into MATLAB and converted into finite element mesh using im2mesh and MESH2D, which is a simplified version of the mesh generation algorithm presented in \cite{Jiexian,Engwirda2014}.

\section{Diffusion and Tortuosity Modeling}\label{diffusionModeling}

\subsection{Governing Equations and Input Properties}

The governing equation for transient moisture diffusion is given by,
\begin{equation}\label{governing equation}
\begin{split}
\frac{\partial C}{\partial t} +\nabla \cdot J = 0 \\
J = -D\nabla C
\end{split}
\end{equation}

\noindent where, $C$ is the moisture concentration, $J$ is the flux associated with moisture and $D$ is the diffusion coefficient of moisture or diffusvitiy. The total domain $\Omega = \Omega_f + \Omega_m $, where $\Omega_f $ is the fiber domain and $\Omega_m$ is the matrix domain as shown in Figure~\ref{schematic}. The exterior boundary of this domain, $\Gamma = \Gamma_1 \cup \Gamma_2 \cup \Gamma_3 \cup \Gamma_4$. Equation~\ref{governing equation} cannot be directly applied to composites because of material heterogeneity. Therefore, a normalized concentration $\phi$ term is used for the diffusion analysis in order to overcome the discontinuities at multi-material interfaces. The normalized concentration and corresponding flux are defined in Equation~\ref{normalizedC}.

\begin{equation}\label{normalizedC}
\begin{split}
    \phi = C/S \\
    J = -D \bigg[S\frac{\partial \phi}{\partial x}+\phi\frac{\partial S}{\partial x}\bigg]
    \end{split}
\end{equation}

where, $S$ is the solubility.

\begin{equation}
\begin{split}
    \phi = 0\quad\Rightarrow \text{Dry}\:\text{state}\\
    \phi = 1\quad\Rightarrow \text{Wet}\:\text{state}\\
\end{split}
\end{equation}
This approach of normalized concentration can be verified using interfacial chemical equilibrium law, where a wetness fraction (normalized concentration) function is introduced in the paper by Wong et al.~\cite{Wong1998}

\begin{figure}[H]
\centering
\subfigure[]{
\includegraphics[width=0.35\textwidth]{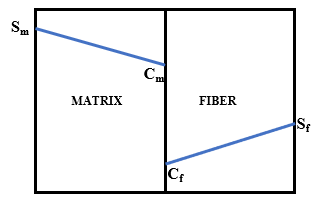}\label{unsaturated_discontinuous}
}
\hspace{0.5in}
\centering
\subfigure[]{
\includegraphics[width=0.35\textwidth]{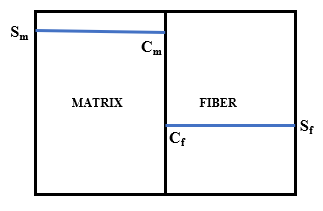}\label{saturated_discontinuous}
}
\caption{Heterogeneous medium with discontinuous concentration at the fiber-matrix interface: (a) Unsaturated (b) Saturated}\label{discontinuous}
\end{figure}

Here, we schematically explain the need for a normalized concentration approach for composites. Consider a bi-material domain as shown in Figure~\ref{discontinuous} which is representative of a FRPC material with different solubility limits for fiber and matrix. Saturation concentration of fiber is not equal to that of the matrix, which introduces a discontinuity at the interface. Figure~\ref{discontinuous} shows a schematic of the concentration in the fiber and matrix regions upon solving the original Fick’s law with concentration is the variable. There exists a discontinuity at the  interface between the two materials. Here, $ S_m $ and $S_f$ are the solubility limits (saturation concentration) of the matrix and fiber, respectively, and $C_m$ and $C_f$ are the moisture concentration of matrix and fiber, respectively. Due to this discontinuity, it is not possible to solve for the concentrations using traditional finite element method.


\begin{figure}[H]
\centering
\subfigure[]{
\includegraphics[width=0.35\textwidth]{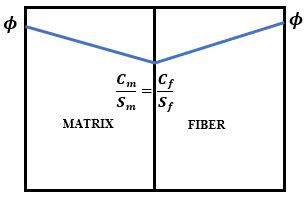}\label{unsaturated_continuous}
}
\hspace{0.5in}
\centering
\subfigure[]{
\includegraphics[width=0.35\textwidth]{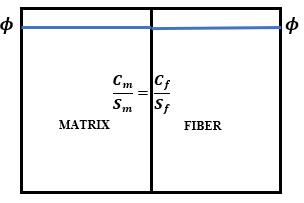}\label{saturated_continuous}
}
\caption{Heterogeneous medium with normalized concentration approach at fiber-matrix interface: (a) Unsaturated and (b) Saturated}\label{continuous}
\end{figure}

Upon normalizing the concentrations with their respective solubilities, the discontinuity at the interface is removed and $\phi$ becomes continuous according to the Nernst partition rule~\cite{Liu2016}. Figure~\ref{continuous} represents the normalized concentration approach, where,
\begin{equation}
    \phi = \frac{C_m}{S_m} = \frac{C_f}{S_f}\quad\Rightarrow \text{Constant}
\end{equation}

The boundary conditions for the 2D micromechanical model are: 
\begin{equation}\label{BC}
\begin{split}
J = 0\quad \text{on}\quad \Gamma_2\cup\Gamma_3\cup\Gamma_4 \quad \Rightarrow\text{no} \: \text{flux} \: \text{boundary} \: \text{condition}\\
\phi = 1 \quad\ \text{on}\quad \Gamma_1 \quad \Rightarrow \text{normalized} \: \text{concentration} \: \text{boundary} \: \text{condition}
\end{split}
\end{equation}

The transient diffusion governing equations described in Equations~\ref{governing equation} and \ref{BC} are solved using finite element method to determine effective diffusivity of moisture in micromechanical models. Quadratic triangular elements (DC2D6) within ABAQUS are used for mesh generation, and a convergence study is performed to determine the optimal mesh. The material properties for both fiber and matrix (resin) are given in Table~\ref{fmproperties}. Here, the diffusivity for fibers is considered to be 0 since the diffusivity of fibers is very negligible compared to that of the matrix \cite{Shen1976}. This assumption is valid for Carbon and Glass fibers, whereas invalid for Kevlar fibers as they absorb significant amount of moisture. Material properties of Epoxy Resin (LY 556) with HT 972 Hardener were obtained from ~\cite{Dattaguru1986}.

\begin{table}[h!]
  \centering
\begin{tabular}{|c|c|c|} 
\hline
Properties & Fiber & Matrix \\
\hline
Diffusivity ($mm^2/s$)& 0 & 1.637 x 10$^{-7}$\\
\hline
Solubility & 1 x 10$^{-36}$ & 2.1\\
\hline
\end{tabular}
\caption{Material properties of fiber and matrix}\label{fmproperties}
\end{table}

\subsection{Effective Diffusivity Calculation}\label{se:diffusivity calculation}
For calculating the coefficient of diffusion or moisture diffusivity for FRPCs, we used the theory proposed by Shen and Springer~\cite{Shen1976} that predicts the coefficient of diffusion under constant temperature and moisture content. They developed an equation that relates moisture content in composites as a function of time with directional diffusivities, geometrical properties and saturation concentration. 
The percentage moisture gain in composites is given by,
\begin{equation}
    M(t) = \frac{W(t)-W_d}{W_d}*100
\end{equation}
where, $W_d$ is the  weight of a dry specimen after manufacturing, that is, at $t=0$. $W(t)$ is the weight of the specimen at a later time instant $t$. 
\begin{figure}[H]
\centering
\subfigure[]{
\includegraphics[width=7cm]{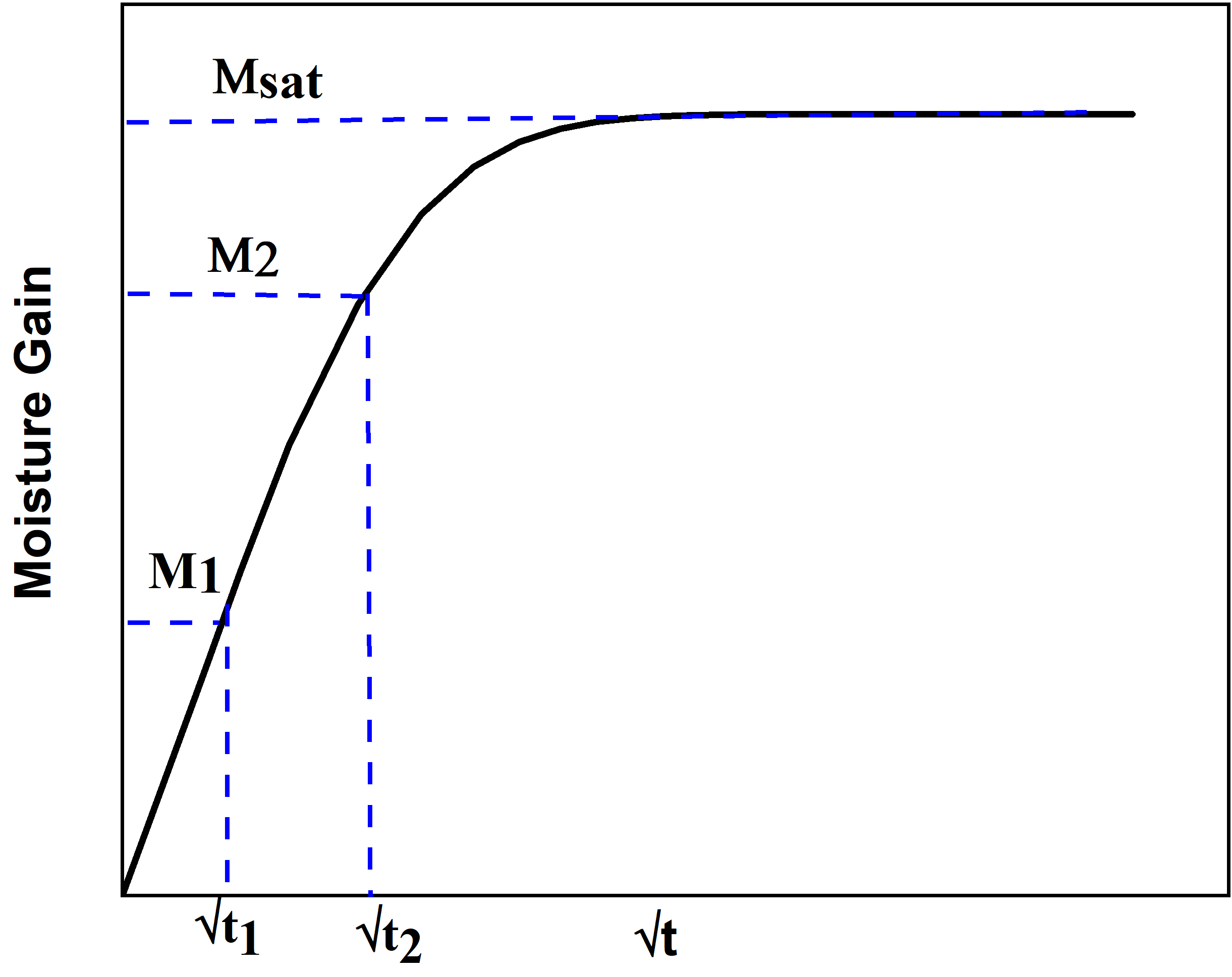} \label{diff_calculation}
}
\caption{Schematic of moisture gain vs time response used for calculating effective diffusivity. Slope of the initial linear region and the plateau moisture saturation value is used in Equation~\ref{D_eff}.}
\end{figure}

Moisture gain as a function of time is given by, 
\begin{equation}\label{moisture_gain_3d}
M(t)= \frac{4M_{sat}}{h\sqrt{\pi}}(\sqrt{D_z}+\frac{h}{L}\sqrt{D_x}+\frac{h}{W}\sqrt{D_y})\sqrt{t}
\end{equation}

where, $M_{sat}$ is the moisture concentration at saturation. $D_x,D_y$ and $D_z$ are the moisture diffusivities along x,y and z directions, respectively. $t$ is time, and $h,W$ and $L$ are the thickness, width and length of the composite, respectively. Shen and Springer~\cite{Shen1976} considered a composite exposed to moisture on the top and bottom surfaces perpendicular to the thickness $h$ direction. They also assumed that the dimensions of the composite along the x and y directions to be infinite, which makes the concentration change along the x and y directions to be negligible. In our current study, since the diffusion is considered to occur primarily along thickness $h$ direction, in addition to $h$ being very small compared to the in-plan dimensions, the moisture gain is considered to primarily occur in the z direction. Hence, the terms $h/L$ and $h/W$ are neglected from Equation~\ref{moisture_gain_3d} making $D_z=D_{eff}$. Equation~\ref{moisture_gain_3d} is rewritten to relate the effective diffusivity $D_{eff}$ with moisture gain as:

\begin{equation}
M(t)= \frac{4M_{sat}}{h}\sqrt{\frac{t}{\pi}}\sqrt{D_{eff}}
\end{equation}

Considering time instants $t=t_1$ and $t=t_2$ in the initial linear region of the moisture gain response, we pick their corresponding moisture gains $M_1$ and $M_2$. Hence, we have,

\begin{equation}
\begin{split}
M_1=M(t=t_1)= \frac{4M_{sat}}{h}\sqrt{\frac{t_1}{\pi}}\sqrt{D_{eff}}\\
M_2=M(t=t_2)= \frac{4M_{sat}}{h}\sqrt{\frac{t_2}{\pi}}\sqrt{D_{eff}}
\end{split}
\end{equation}

Subtracting the above two equations and rearranging gives:

\begin{equation}\label{D_eff}
    D_{eff}=\pi\left(\frac{h}{4M_{sat}}\right)^2\left(\frac{M_2- M_1}{\sqrt{t_2}-\sqrt{t_1}}\right)^2
\end{equation}

Diffusivity is numerically calculated by first performing transient moisture diffusion analysis on micromechanical models. Moisture gain percentage the mass concentration at the centroid of each element. These values are averaged over the entire domain at several time instants to obtain the moisture gained by the composite over a time period until it attains moisture saturation concentration. Using the initial slope of the moisture gain curve, thickness along the flow direction and $M_{sat}$, the effective diffusivity $D_{eff}$ is calculated using Equation~\ref{D_eff}. An empirical method for calculating effective diffusivity was also proposed by Shen and Springer~\cite{Shen1976} based on the similarities between heat conduction and mass diffusion. Here, they modified the analytical equation proposed by Springer and Tsai~\cite{Springer1967} for heat conduction to mass diffusion analysis. The transverse diffusivity is given by,
\begin{equation}
    D_{22}=(1-2\sqrt{\frac{V_f}{\pi}})D_m
\end{equation}

where $D_m$ and $V_f$ are the diffusion coefficient of resin and fiber volume fraction, respectively.

\subsection{Tortuosity Calculation}\label{tortuosity_calculation}
Tortuosity has been predominantly used in electrochemical devices with highly complex electrode microstructure, which results in very tortuous electrolyte flow that plays an important role in ionic conductivity~\cite{Suthar2015,Froboese2019,Forouzan2017}. Tortuosity in a porous media is a significant parameter that characterizes diffusive flow within hydraulic transport, mass transport, electric, heat  or ionic conduction. Several porosity-tortuosity associations have previously been published based on experiments, and a detailed review is given by Shen~\cite{Shen2007}.

Tortuosity is geometrically defined as the ratio of the length of a convoluted path of a diffusive flow in a porous medium to the length of the shortest path in a homogeneous media.
\begin{equation}\label{geometric_tau}
    \tau = \frac{L_e}{L}
\end{equation}
where, $L_e$ is the length of a convoluted path and $L$ is the length of the shortest.

Originally, Epstein~\cite{Epstein1989} coined a tortuosity factor that accounts for both the path length and change in diffusivity in a porous media. In their work~\cite{Epstein1989}, they use tortuosity factor within diffusion theory, where effective diffusivity $D_{eff}$  of a porous medium is calculated based on the bulk diffusion coefficient.
\begin{equation}\label{diff_tort}
    D_{eff}=\frac{\epsilon}{\tau}*D_{bulk}
\end{equation}
where, $D_{eff}$ , $D_{bulk}$, $\epsilon$ and $\tau$ are the effective diffusivity, bulk diffusivity, pore volume fraction and tortuosity, respectively.

In 1935, Bruggeman~\cite{VonD.A.G.Bruggeman1935} proposed a Porosity-Tortuosity empirical relationship shown in Equation~\ref{brugg}, which is commonly recognized in the field of porous media flow and is often used to derive materials' effective properties. They assumed that the heterogeneous media consisted of only spherical particles. 
\begin{equation}\label{brugg}
    \tau=\frac{1}{\epsilon^{\frac{1}{2}}}
\end{equation}

Both geometrical and flux based algorithms have been developed by researchers for determining the tortuosity factor. In geometry based algorithms\cite{Vivet2011,Cooper2014}, tortuosity is calculated as defined in Equation~\ref{geometric_tau} using the shortest pathway for which the pore centroid method and the fast marching method are commonly used. Whereas, in flux-based algorithms\cite{Izzo2008,Nanjundappa2013,Wilson2006}, the tortuosity factor is calculated by considering diffusion behaviour. In addition, flux based algorithms can be subdivided into voxel-based and mesh-based algorithms. We used a mesh-based approach~\cite{Wilson2006} in this paper, where we perform finite element analysis to determine flux in heterogeneous media as described next. Mesh sensitivity analysis is performed to reduce the effect of the finite element mesh on the tortuosity calculation. 
\begin{figure}[H]
\centering
\subfigure[]{
\includegraphics[width=5cm]{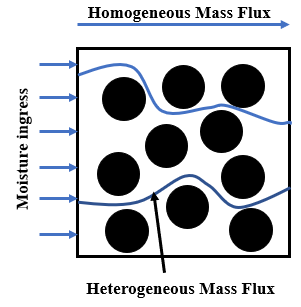} \label{tau_schematic}
}
\hspace{0.5in}
\centering
\subfigure[]{
\includegraphics[width=4.5cm]{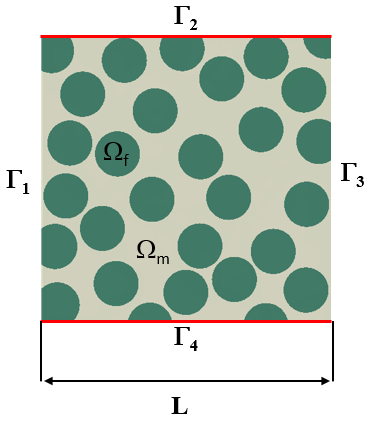}\label{tau_governing}
}
\caption{(a) Schematic representation of diffusive flow in heterogeneous medium with impermeable inclusions; (b) Micromechanical model with regions and boundaries defined}
\end{figure}

For calculating tortuosity factor, we run steady state diffusion analyses on micromechanical models with different fiber volume fractions as well as a model with matrix only. The governing equation and the boundary conditions are as follows:
\begin{equation}\label{laplace}
\begin{split}
    \nabla J = -D\nabla^2 C=0\\
    C = 1\quad on \quad \Gamma_1\\
    C=0 \quad on \quad \Gamma_3\\
    J =0\quad on \quad \Gamma_2\quad and \quad \Gamma_4
    \end{split}
\end{equation}

Since tortuosity depends on the flow path, we obtained a tortuosity factor in the direction in which the moisture ingression occurs. By solving the Laplace equation shown in Equation~\ref{laplace} using finite element method, volume average of flux $J$ in the heterogeneous medium is determined. The volume average of flux $J_{MF}$ in a homogeneous medium free of barriers (in our case, ﬁbers) is also calculated. The tortuosity factor is determined using the expression shown in Equation~\ref{tauFactor}, which is the same as Equation~\ref{diff_tort}.

\begin{equation} \label{tauFactor}
\begin{split}
    -\frac{\epsilon}{\tau}\nabla C_{MF} = -\frac{1}{V}\int\int\int_V\nabla C \: dV   \Rightarrow -\frac{\epsilon}{\tau} J_{MF} = -\frac{1}{V}\int\int\int_V J \: dV \\[10pt]
    \Rightarrow  {\tau} = \frac{J_{MF}}{\frac{1}{V}\int\int\int_V J \: dV} \: \epsilon \\[10pt]
    \Rightarrow \tau = \left(\frac{\text{Volume averaged flux of homogeneous model}}{\text{Volume averaged flux of heterogeneous model}}\right)\\ \times \text{(Volume fraction of moisture absorbing media})
    \end{split}
\end{equation}

Here, $V$ is the volume of the heterogeneous domain, $\tau$ is the tortuosity factor and $\epsilon$ is generally termed as porosity or in our case the volume fraction of domain that is available for the movement of flux. In this paper, since we consider that only the matrix absorbs moisture, $\epsilon$ can be defined as 1-$V_f$ where $V_f$ is the fiber volume fraction.

\section{Results and Discussion}\label{se:Results}

\subsection{Effect of Fiber Volume Fraction and Spatial Distribution on Tortuosity and Moisture Diffusivity}\label{sse:effect_of_vf}

\begin{figure}[H]
 \centering
  \includegraphics[width=12cm]{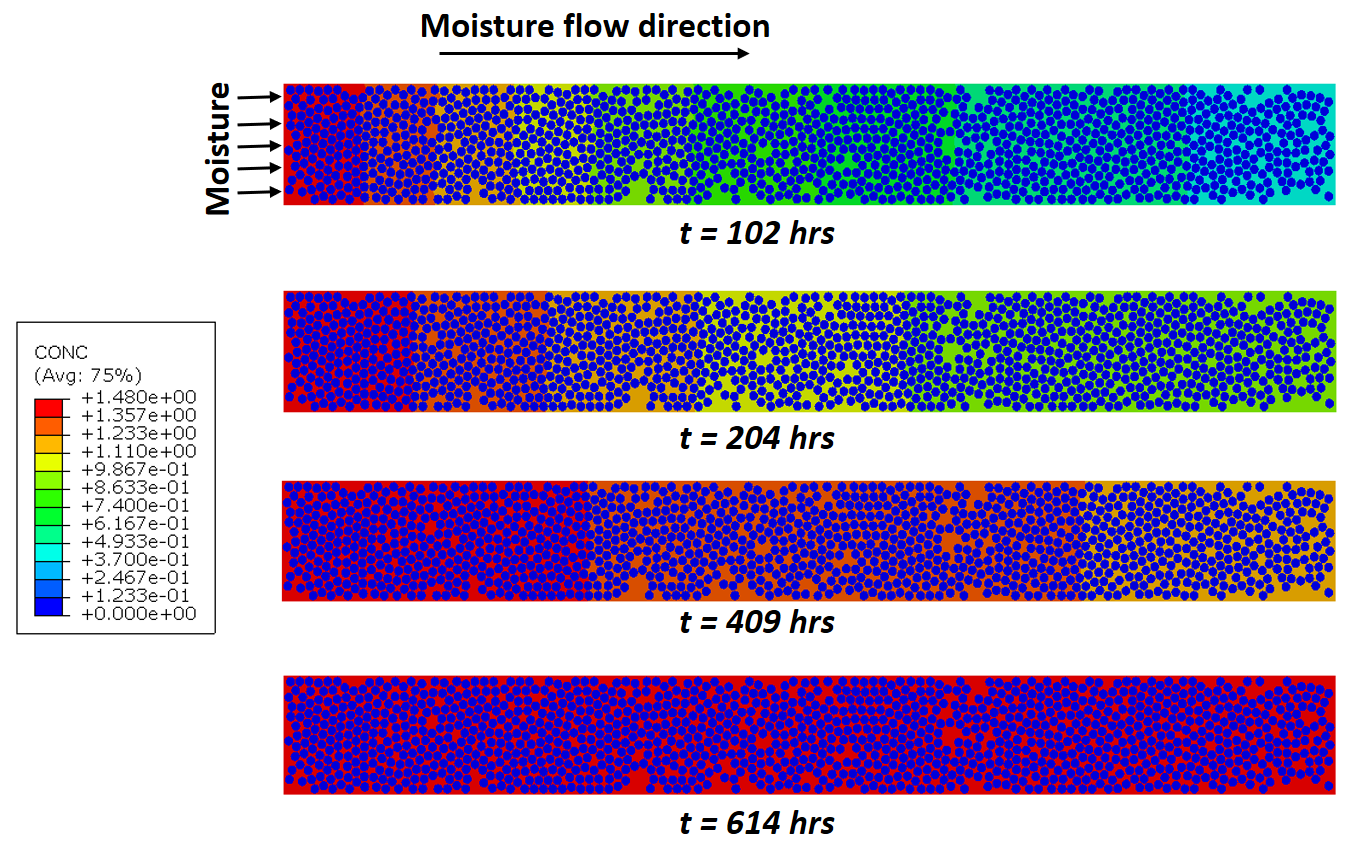}
  \caption{Validation with Vaddadi et al. \mbox{\cite{Vaddadi2003}} - Transient moisture ingression (concentration) contours in carbon fiber reinforced polymer composite at different time intervals.}\label{validation_contour}
\end{figure}

\begin{figure}[H]
 \centering
  \includegraphics[width=7cm]{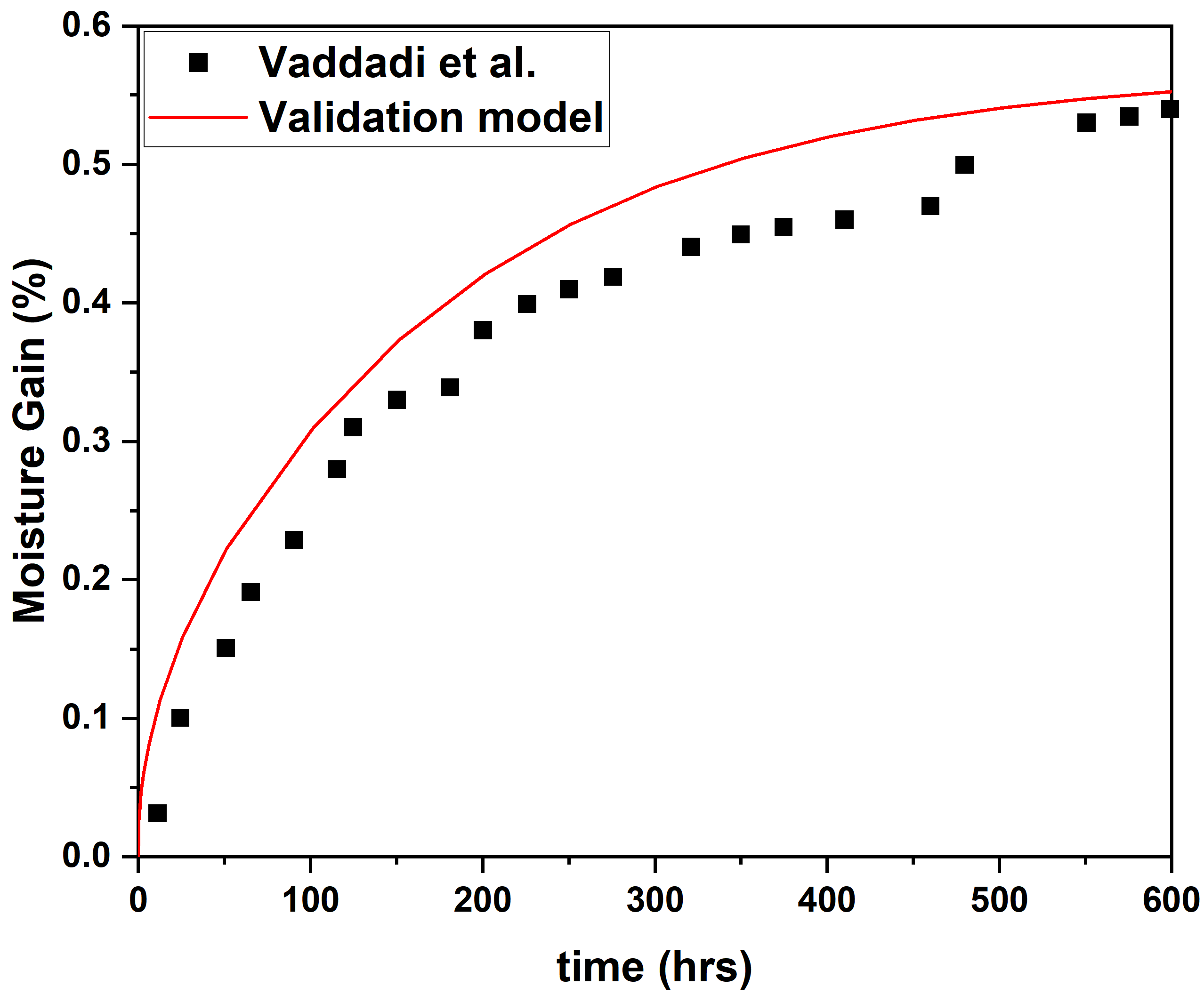}
  \caption{Validation of our numerical model with experimental results~\mbox{\cite{Vaddadi2003}} - Comparison of moisture gain (\%) versus time} \label{validation_moisture_gain} 
\end{figure}

Our micromechanical model was first validated against experimental results from \mbox{Vaddadi et al.~\cite{Vaddadi2003}} by comparing the rate of moisture gain and moisture saturation. These 2D micromechanical model comprised of carbon fibers and epoxy matrix system. This numerical model for validation considers moisture ingression for the case of temperature = 85\textdegree C and relative humidity of 85\%. Carbon fibers are considered to be impermeable to moisture. Material properties for the matrix are considered as provided by Vaddadi et al. \mbox{\cite{Vaddadi2003}} 
The validation model is 600\textmu m x 64 \textmu m and consists of 1200 fibers of 5 \textmu m diameter. The moisture ingression happens at the left edge of the sample with no flux boundary condition at all other edges as shown in \mbox{Figure~\ref{validation_contour}}. \mbox{Figure~\ref{validation_moisture_gain}} shows the comparison of the rate of moisture gain and moisture saturation between the experimental and our numerical model. We could see small difference in rate of weight gain between the experimental and our numerical validation model which is mainly attributed to variation in the microstructural distribution between the two.

\begin{figure}[H]
 \centering
  \includegraphics[width=7cm]{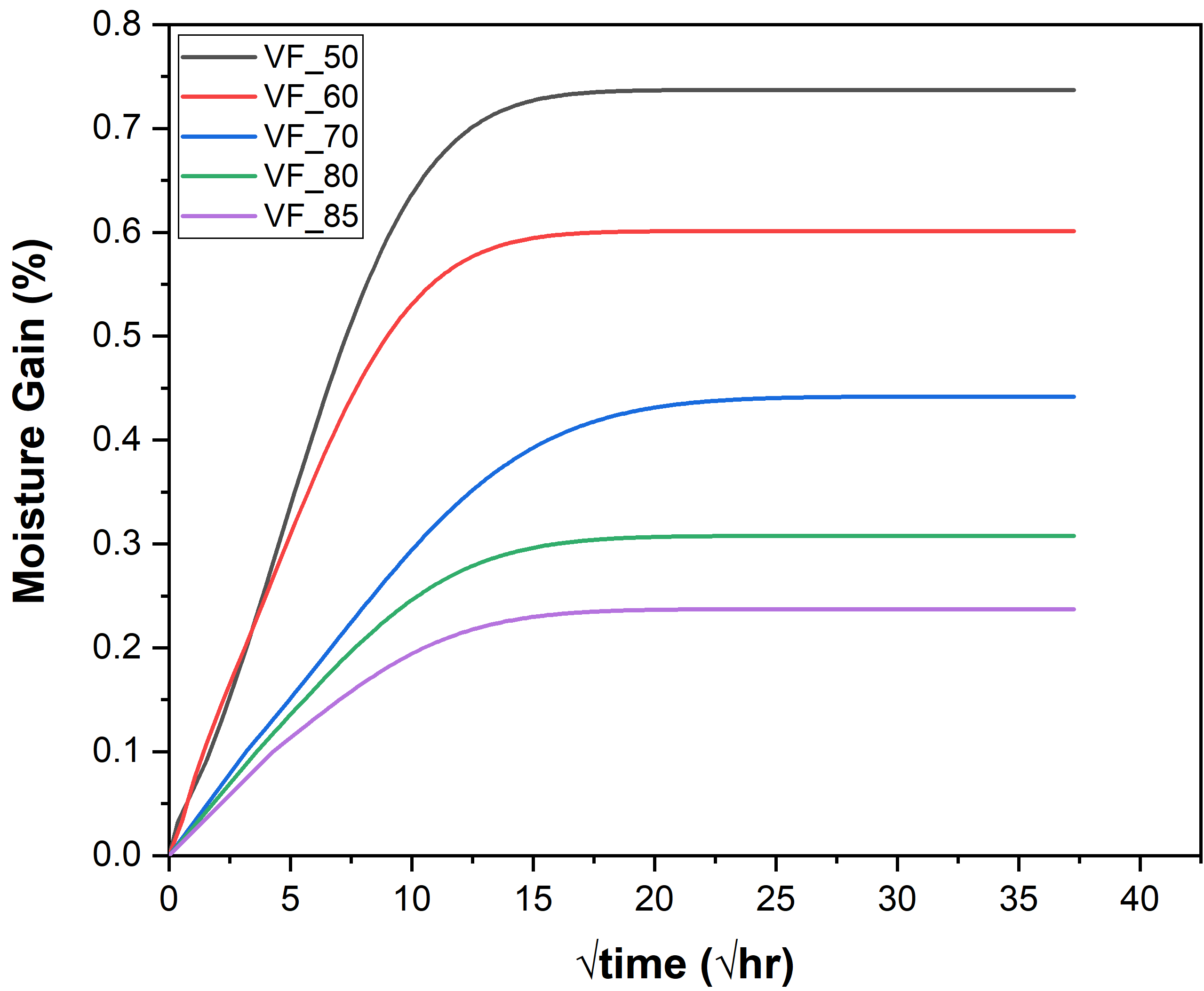}
  \caption{Moisture gain response for different volume fractions}\label{weight_gain_volume_fraction}
\end{figure}

Next, 2D micromechanical models with different fiber spatial arrangements like square, hexagonal and random arrays with a range of fiber volume fractions of 50, 55, 60, 65 and 70\text{\%} were considered for calculating diffusion and tortuosity. The geometry, material properties, governing equation and boundary conditions were mentioned previously in Section~\ref{diffusionModeling}. In this model, one edge of the domain was exposed to moisture, while all the other edges were considered impervious to moisture. The average concentration (weight gain percentage) of moisture in the entire domain was recorded at different times from which the weight gain curves for different volume fractions were plotted. Using these weight gain plots, the effective diffusivity of composites was calculated using the method described in Section~\ref{se:diffusivity calculation}. As shown in Figure~\ref{weight_gain_volume_fraction}, the moisture saturation concentration changes with volume fraction of fibers.

\begin{figure}[H]
\centering
\subfigure[]{
\includegraphics[width=7cm]{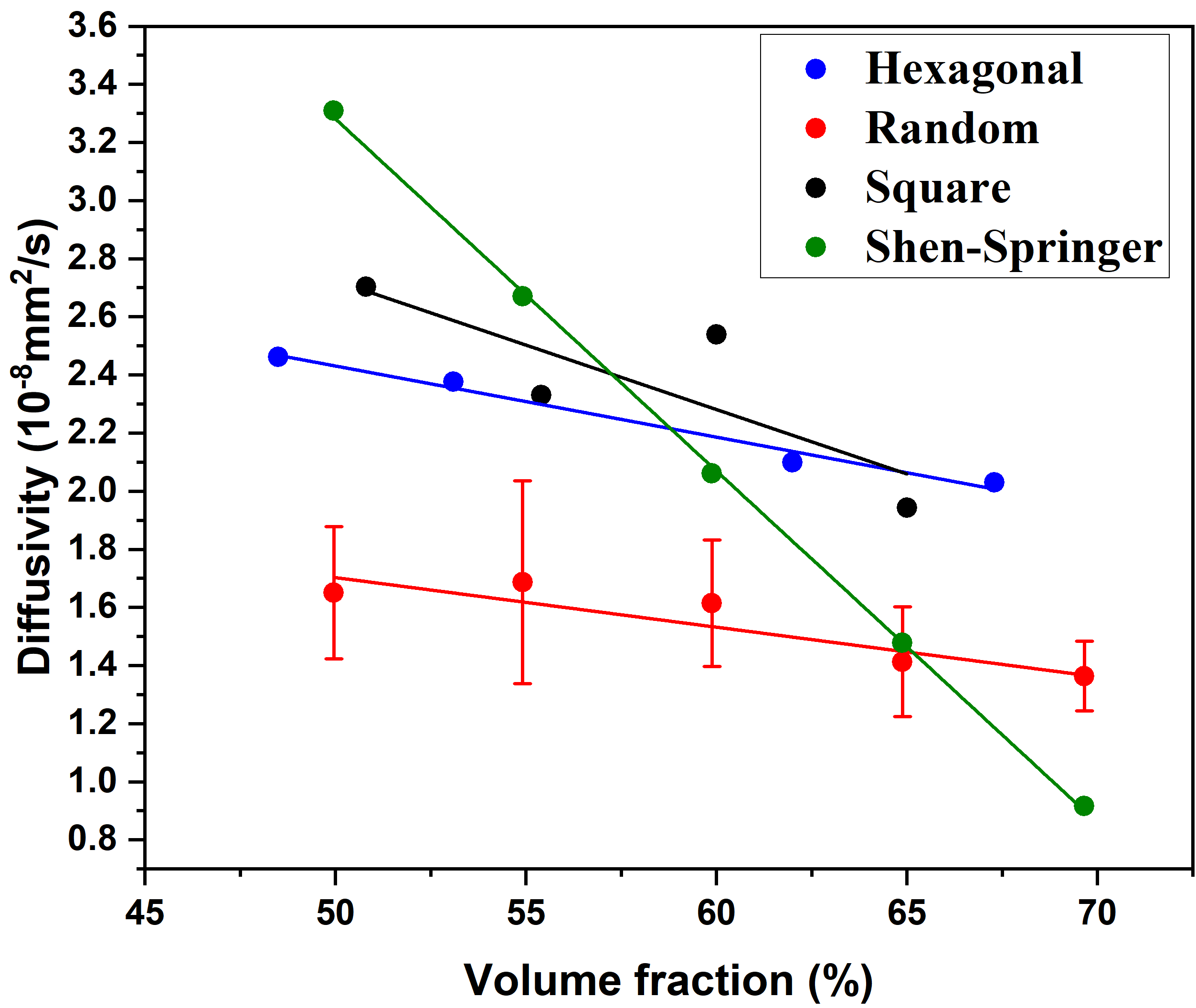}\label{Diffusivity_plot}
}
\hspace{0.2in}
\centering
\subfigure[]{
\includegraphics[width=7cm]{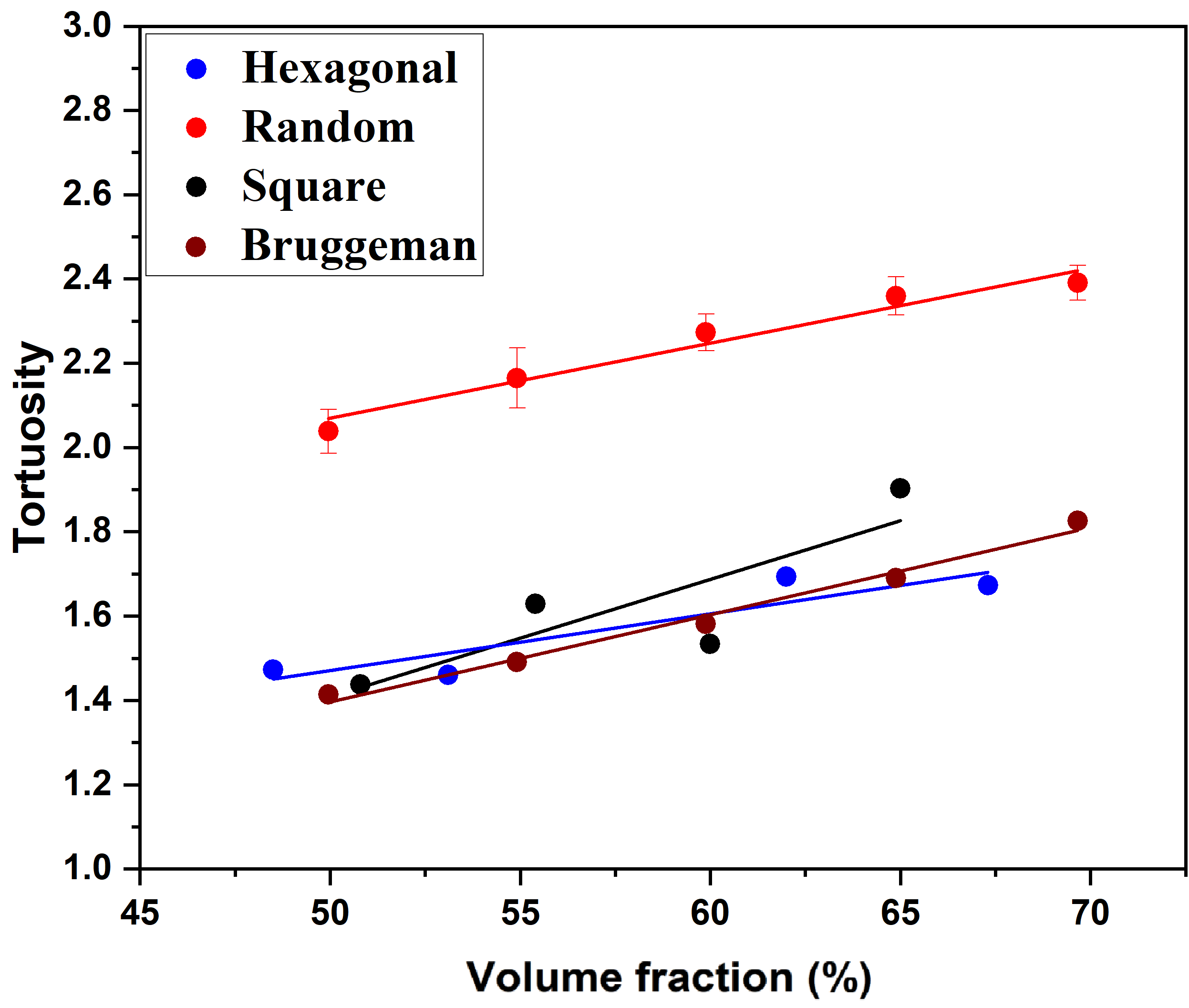} \label{tau_plot}
}
\caption{(a) Diffusivity and (b) Tortuosity versus fiber volume fraction with linear trend lines}\label{fig:diff_tau}
\end{figure}

The effective diffusivity is plotted against volume fraction in Figure~\ref{Diffusivity_plot} for various fiber arrangements and an analytical method, where the effective diffusivity decreases with an increase in volume fraction for all models. Among all fiber spatial arrangements considered, the effective diffusivity of random fiber model is the lowest. One of the key reasons for this behavior is the disorderliness of fiber arrangement in this model. As mentioned before, we considered the Carbon fibers to be impermeable \cite{Dattaguru1986,Whitcomb2002,Shen1976} in our models, as the moisture absorption by fibers is insignificant compared to matrix (resin). The fibers being impermeable to moisture, act as barriers that hinder the flow of moisture into the composites, thereby, increasing the time to reach saturation. Due to this behavior, the effective diffusivity is lower for the random fiber distribution model compared to those with regular patterns, like square and hexagonal array. 

Tortuosity factor (Figure~\ref{tau_plot}) is numerically calculated for different fiber spatial distributions and volume fractions for quantitatively depicting the disorderliness of fibers that cause tortuous diffusive moisture flow pathways into the composite. The tortuosity is also calculated using the Bruggeman relationship given in Equation~\ref{brugg}. We observe that these values lie between the numerically calculated tortuosity values from square and hexagonal array models. In general, we observe that tortuosity is inversely proportional to effective diffusivity for all cases as seen from Figure~\ref{Diffusivity_plot} and Figure~\ref{tau_plot}. The tortuosity for randomly distributed fibers is higher compared to that for square and hexagonal array models, which is attributed to fiber distribution. In a regular array, the diffusive flow of moisture is less tortuous compared to that in a random array. An increase in tortuosity decreases the effective diffusivity for each fiber distribution, which is evident from Figures~\ref{Diffusivity_plot} and ~\ref{tau_plot}.

\begin{figure}[h!]
 \centering
  \includegraphics[width=7cm]{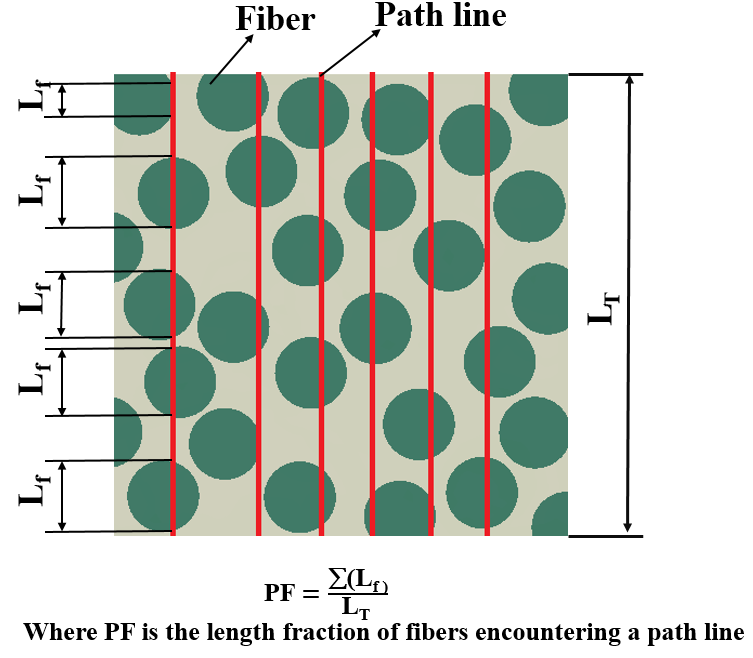}
  \caption{Pictorial representation of perturbation fraction calculation (Equation~\ref{eq:perturbFactor}) using information about fiber spatial arrangement}\label{Perturb_calc}
\end{figure}

\begin{figure}[h!]
\centering
\subfigure[]{
\includegraphics[width=7cm]{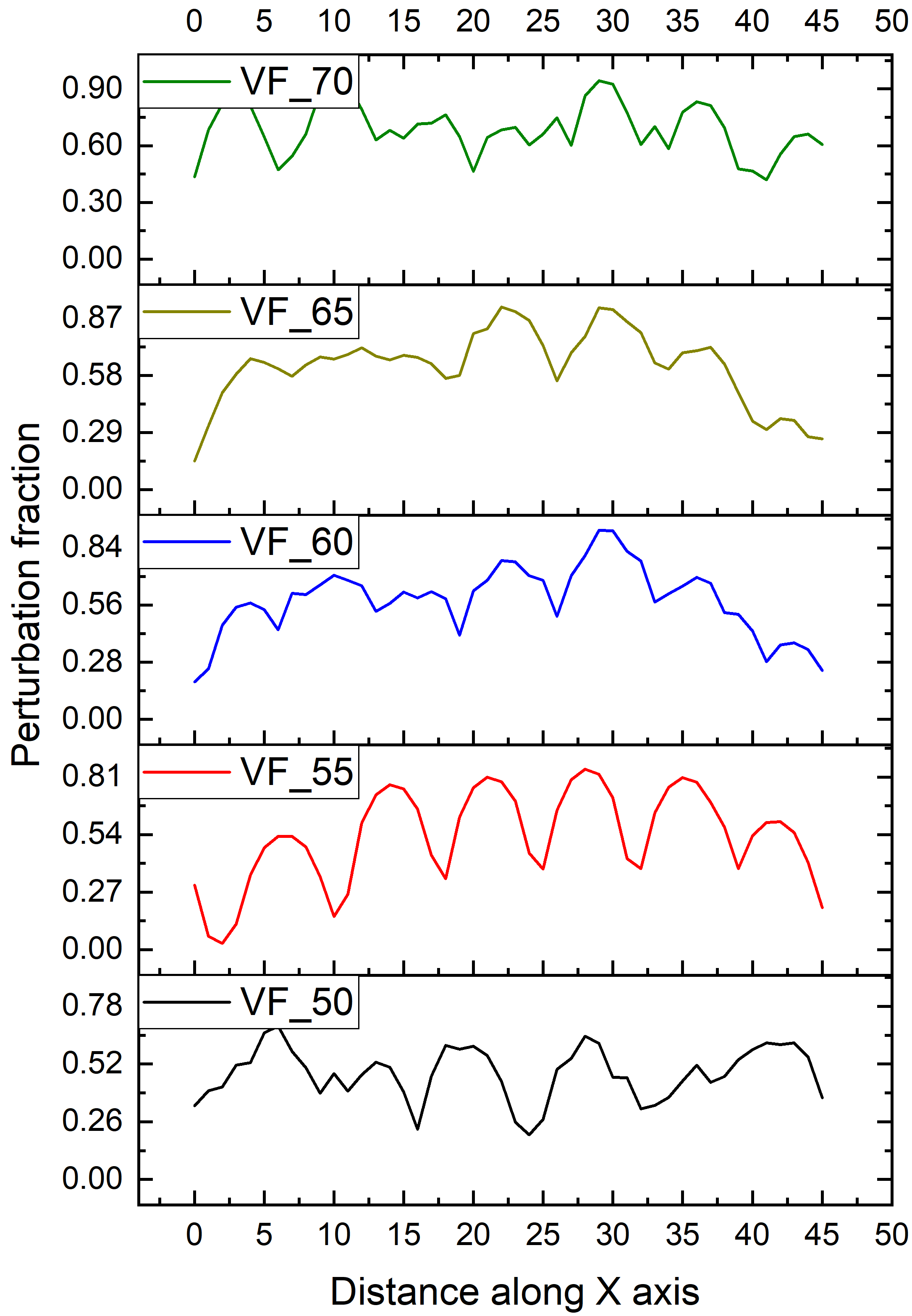} \label{Random_perturb}
}
\centering
\subfigure[]{
\includegraphics[width=7cm]{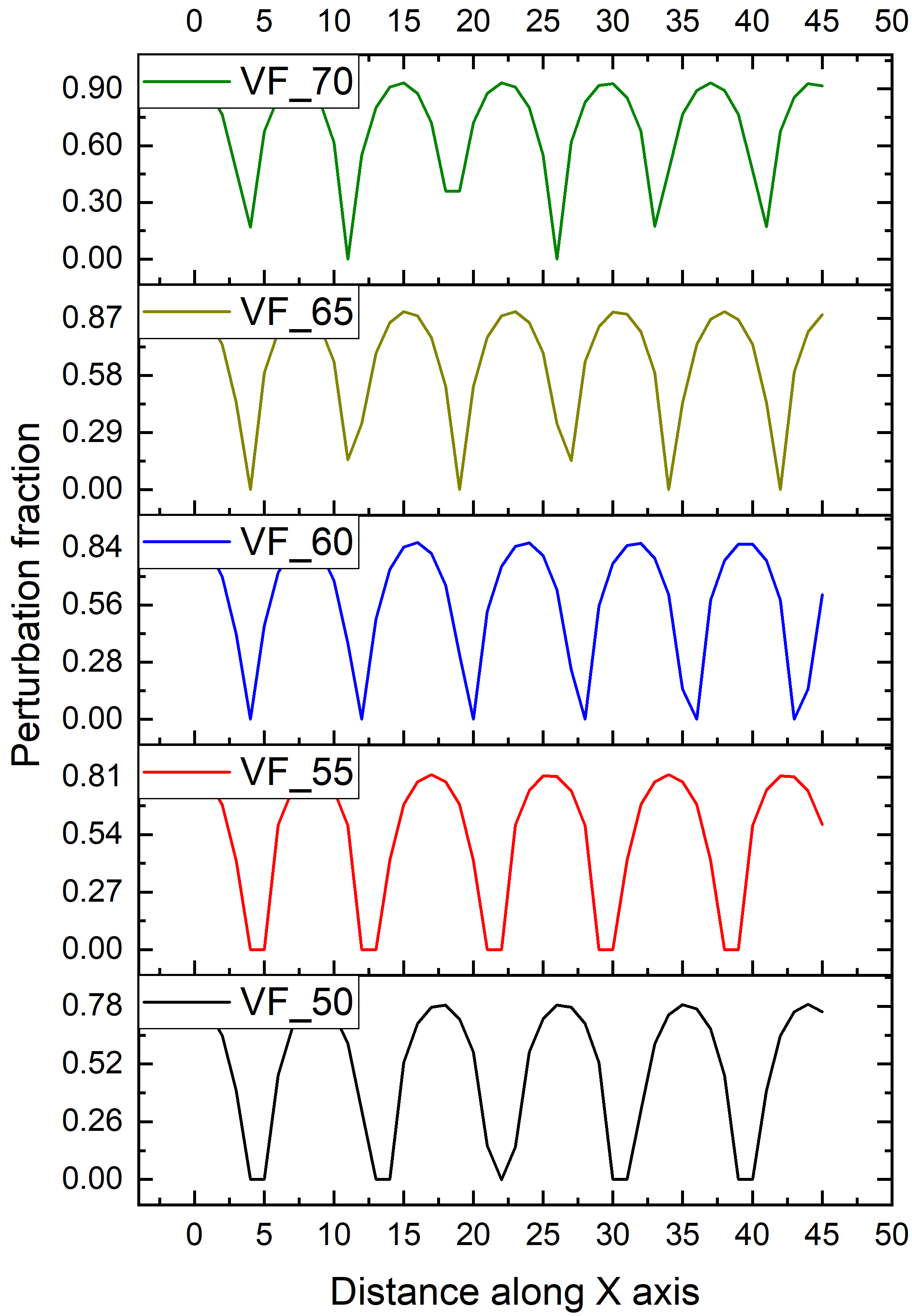}\label{square_perturb}
}
\caption{Perturbation fraction along X axis in: (a) Random array and (b) Square array models}\label{Perturb_plots}
\end{figure}

To further examine the impact of spatial distribution of fibers on tortuosity, we established a length fraction called ``Perturbation Fraction'' that qualitatively reflects fiber disorderliness. As schematically shown in Figure~\ref{Perturb_calc}, numerous path lines are created in a micromechanical model along the height of the domain at regular intervals. Perturbation Fraction (PF) is defined as the sum of the lengths of each path line encountering fibers $L_f$ divided by the total height of the domain $L_T$ (Equation~\ref{eq:perturbFactor}). PF at regular intervals along the x-axis is plotted in Figure~\ref{Random_perturb} and Figure~\ref{square_perturb} for random and square fiber arrangements, respectively, for different fiber volume fractions.
\begin{equation}\label{eq:perturbFactor}
    PF = \frac{\sum (L_f)}{L_T}
\end{equation}

As shown in Figure~\ref{Perturb_plots}, PF is plotted along the length of the whole domain. Essentially, every point on these plots is a PF value along a vertical path line. These PF plots help us to visualize fiber distribution across the entire domain, which is directly associated with the tortuosity of diffusion pathways. We observe that the variation of PF in a square array is periodic across all fiber volume fractions. The lowest PF value is at 0\text{\%}, which means that there are clear channels of matrix for moisture to flow that makes a diffusive pathway less tortuous. In case of random fiber distribution, the variation of PF is random across all fiber volume fractions and the lowest value is 26\text{\%}. This makes the diffusion route more tortuous and corroborates our earlier findings regarding tortuosity in Figure~\ref{tau_plot}.

\subsection{Effect of Fiber Morphology on Tortuosity}\label{sse:effect_of_fiber_shape}
One of the main objectives of this paper is to study real microstructures that are typically a combination of circular fibers and many other shapes that have an impact on moisture diffusivity. To investigate the effect of fiber morphology on tortuosity, we considered various fiber shapes including several types of curved bean fibers, asteroid, square, octagonal, elliptical and circular fibers. We modeled fiber cross-sections with different morphology but fixed cross-sectional area to understand the influence of fiber cross-sectional perimeter (in 2D space) and orientation of fiber shape on tortuosity. We defined a "Normalized Perimeter" term, which is the ratio of the perimeter of a fiber with non-circular shape to the circumference of a circular fiber.

\begin{figure}[h!]
 \centering
  \includegraphics[width=0.5\textwidth]{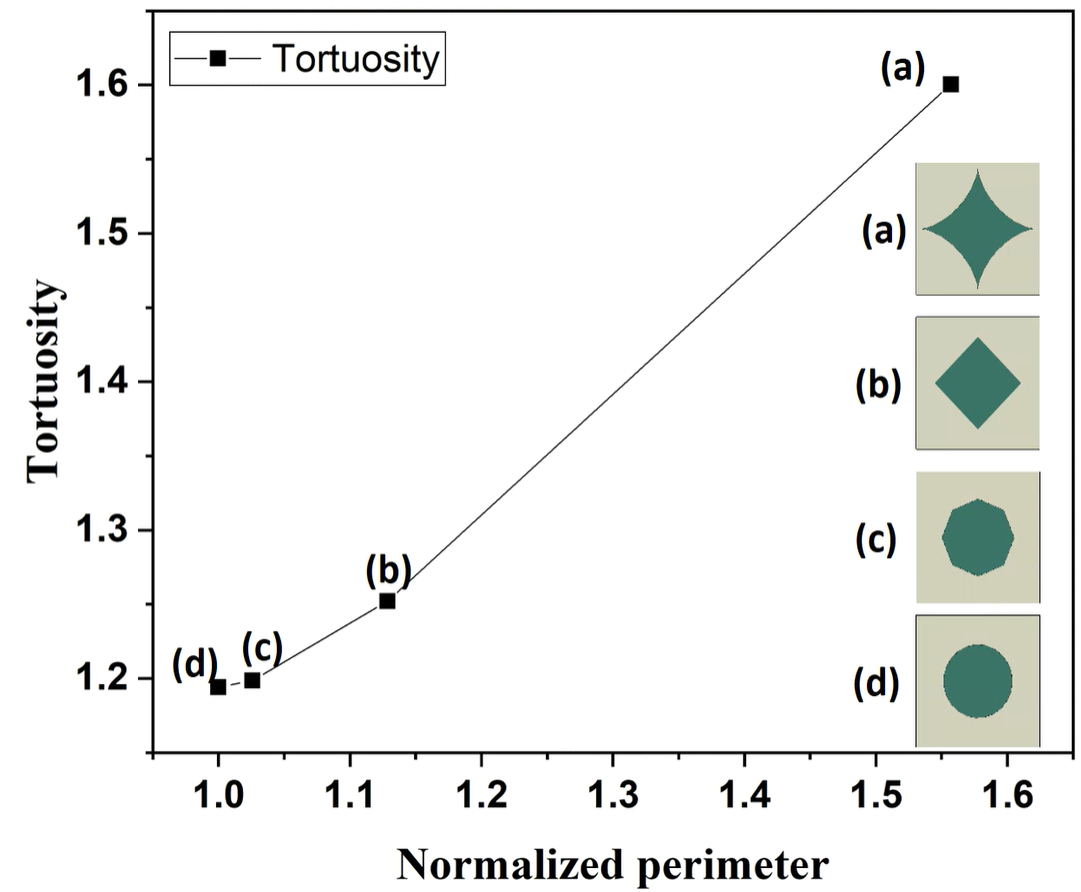}
  \caption{Influence of perimeter of cross-section on tortuosity}\label{Perimeter_tort}
\end{figure}

First, we evaluated the impact of fiber perimeter on tortuosity by considering asteroid, cube, circle, and octagonal fibers embedded in a matrix medium. These shapes were particularly chosen as they are symmetric and do not have directional dependencies. Tortuosity is plotted against normalized perimeter in Figure~\ref{Perimeter_tort} for these different shapes considered. We observe that as the normalized perimeter increases, the tortuosity increases with the highest tortuosity for asteroid, which has the largest perimeter compared to all other shapes considered. Hence, we conclude that larger perimeter has greater hindrance to moisture ingression, thereby, decreasing the moisture diffusivity. The tortuosity of an octagon is marginally higher than that of a circle as its normalized perimeter is close to 1.

\begin{figure}[h!]
\centering
\subfigure[]{
\includegraphics[width=0.35\textwidth]{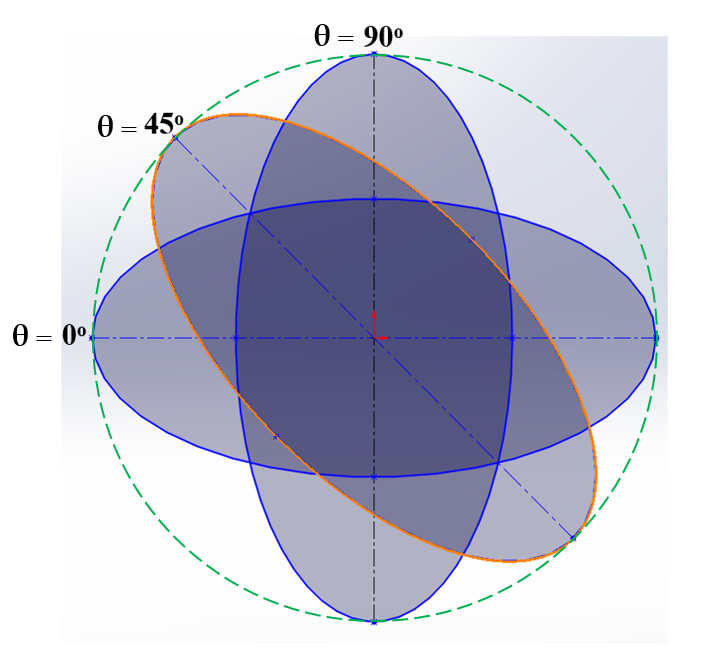} \label{orientation}
}
\hspace{0.2in}
\centering
\subfigure[]{
\includegraphics[width=0.4\textwidth]{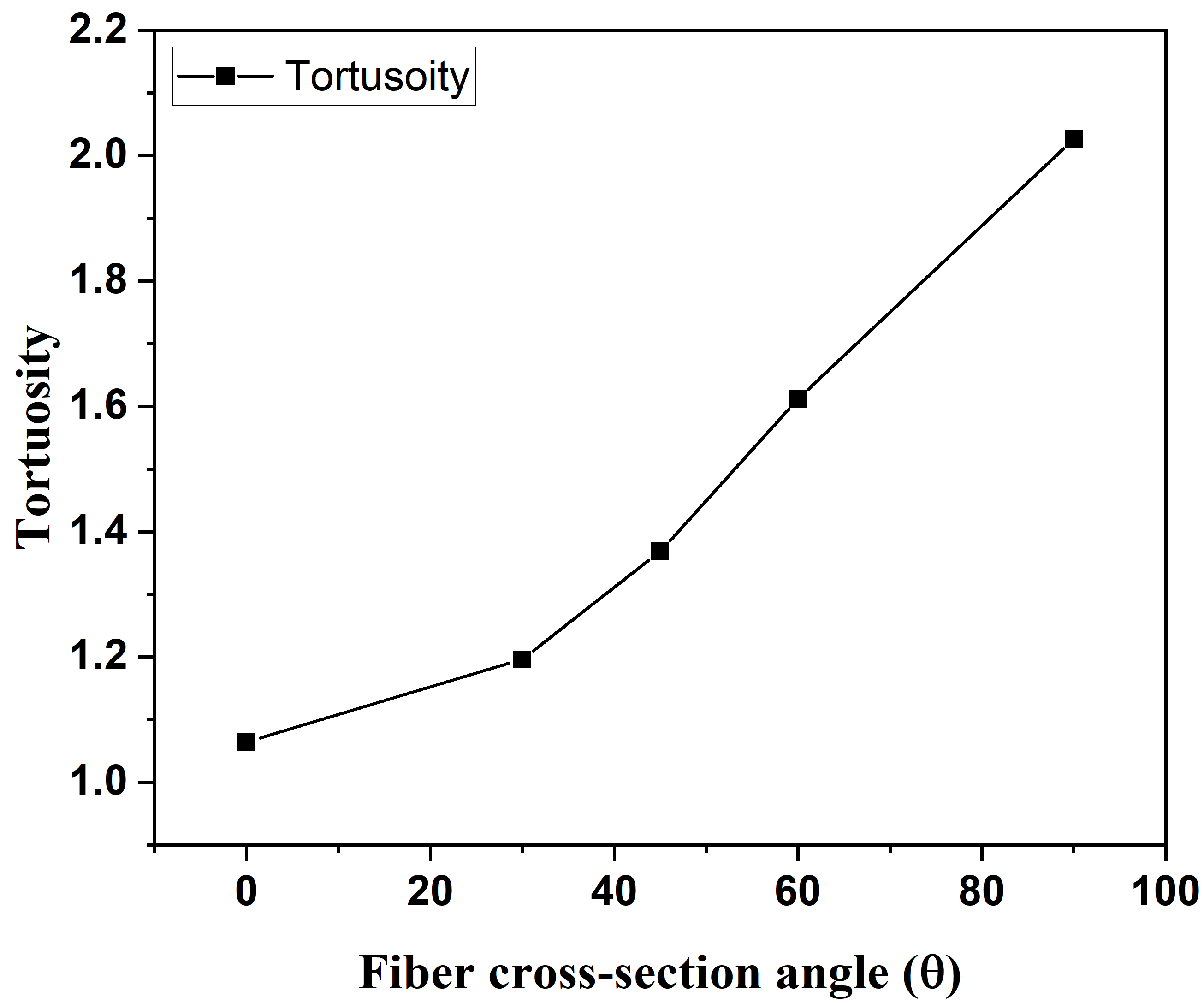}\label{orientation_tortuosity}
}
\caption{(a) Cross-sectional angle of elliptical fiber; (b) Tortuosity vs fiber cross-section angle of ellipse}
\end{figure}

In addition to fiber perimeter, the fiber cross-sectional angle $\theta$, that is the angle that the cross-section makes in its plane, can also have an impact on tortuosity and moisture diffusivity. To understand this effect, we considered elliptical fiber with varying cross-sectional angle from $0^\circ$ to $90^\circ$ as shown in \mbox{Figure~\ref{orientation}}. Tortuosity is calculated for different fiber cross-sectional angles in this range and plotted in \mbox{Figure~\ref{orientation_tortuosity}}. We observe that the tortuosity increases with increase in fiber cross-sectional angle of the elliptical fiber. The main reason for this behavior when moisture flow occurs from left to right is as follows: when the angle is $0\deg$, the distance over which the moisture flow occurs (impact length) is equal to the minor axis or shortest length in an elliptical cross-section. But, as the cross-sectional alignment angle increases, this impact length for moisture flow increases. Thus, the tortuosity is maximum when the fiber cross-sectional angle is along the major axis or longest length of an elliptical cross-section.

\begin{figure}[H]
 \centering
  \includegraphics[width=6cm]{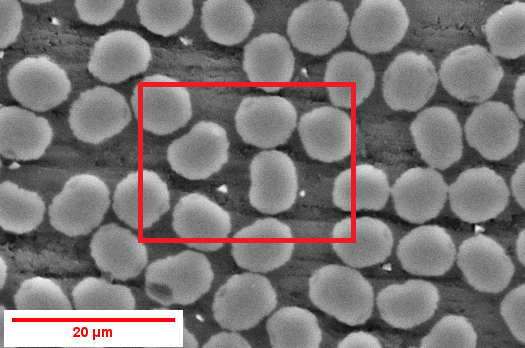}
  \caption{Scanning electron microscope image of  -  dry Carbon Fiber Reinforced polymer microstructure that comprises of different fiber morphology(in the red box)}\label{real_microstructure_morphology}
\end{figure}

\begin{figure}[H]
\centering
\subfigure[]{
\includegraphics[width=7cm]{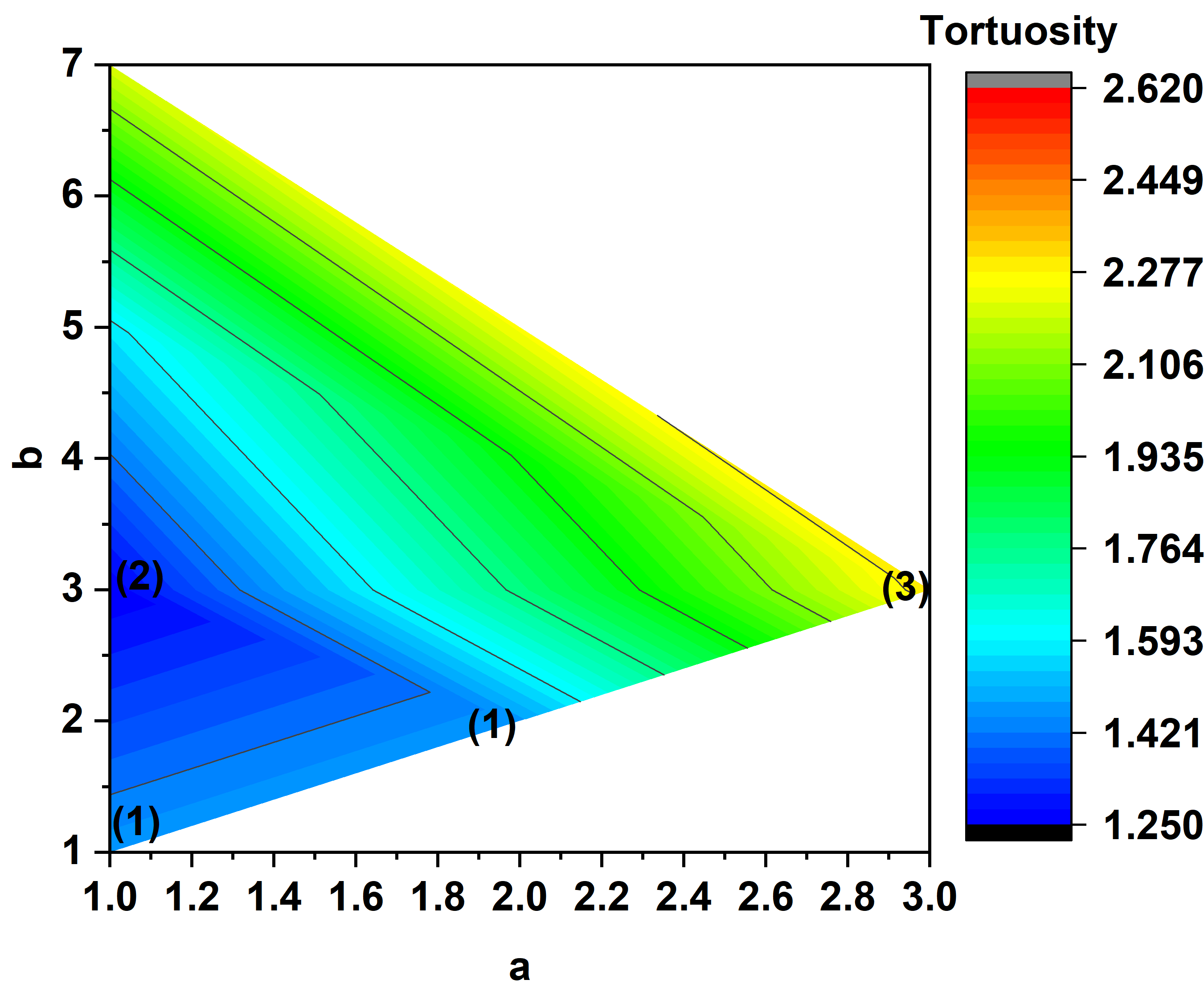} \label{shape_x}
}
\hspace{0.2in}
\centering
\subfigure[]{
\includegraphics[width=7cm]{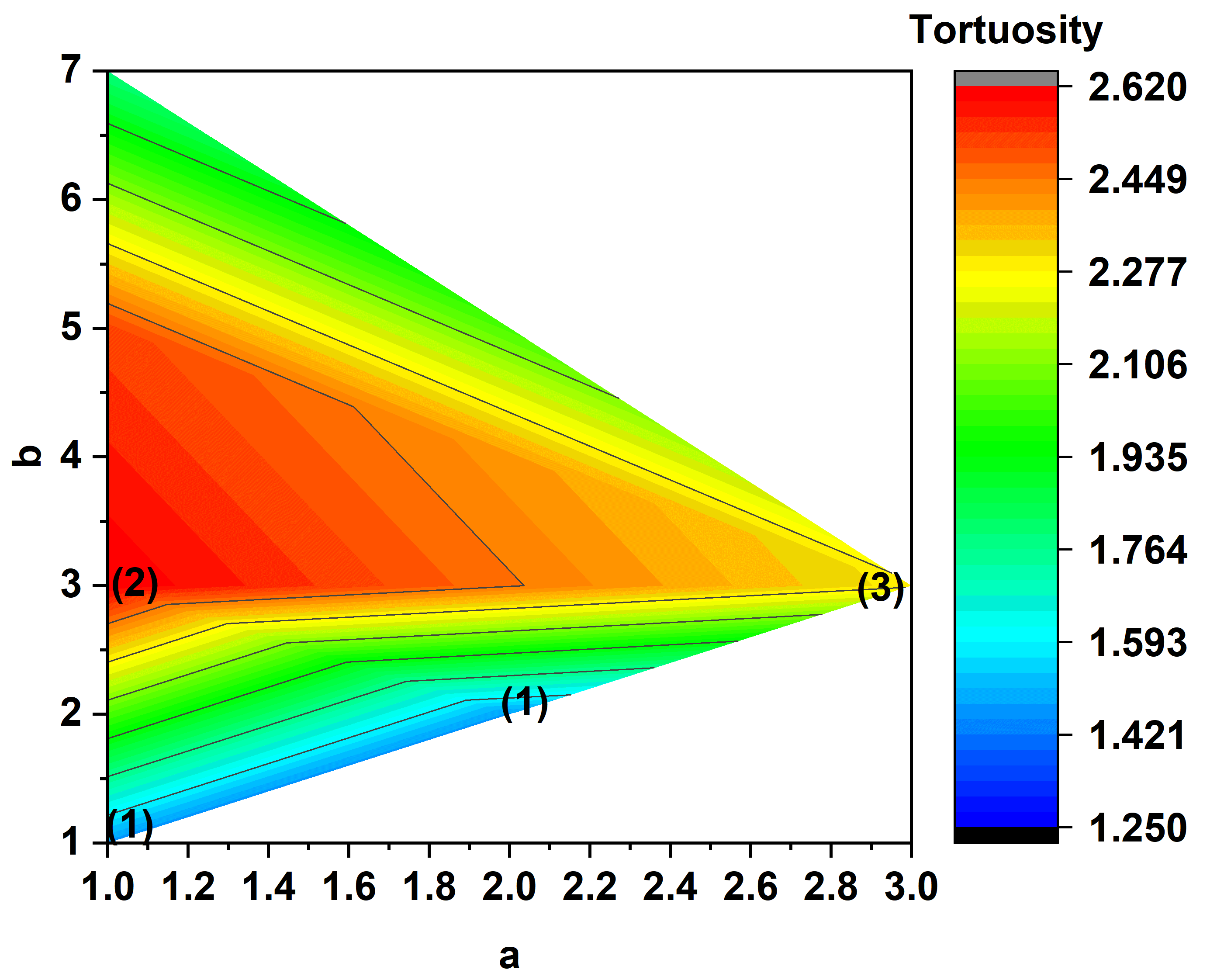}\label{shape_y}
}
\caption{Tortuosity of different bean curves with the variation of the geometrical parameter a and b: (a) diffusive flow along x direction and (b) diffusive flow along y direction}
\end{figure}

Figure~\ref{real_microstructure_morphology} shows the SEM image of the cross-section a real composite at the microscale. This contains carbon fibers with different morphology, and are typically various forms of bean curves. A unit cell with 50\text{\%} fiber volume fraction is considered for numerical modeling, where the cross-sectional area of all the fiber morphologies considered are kept constant. The governing equation and boundary conditions are applied as described previously in Section~\ref{tortuosity_calculation} in order to quantify the tortuosity. Figure~\ref{shape_x} and Figure~\ref{shape_y} show maps of tortuosity values for different shapes, where the moisture flow is considered to occur along the x and y directions, respectively. 

\begin{figure}[H]
    \centering
\subfigure[]{
\includegraphics[width=4cm]{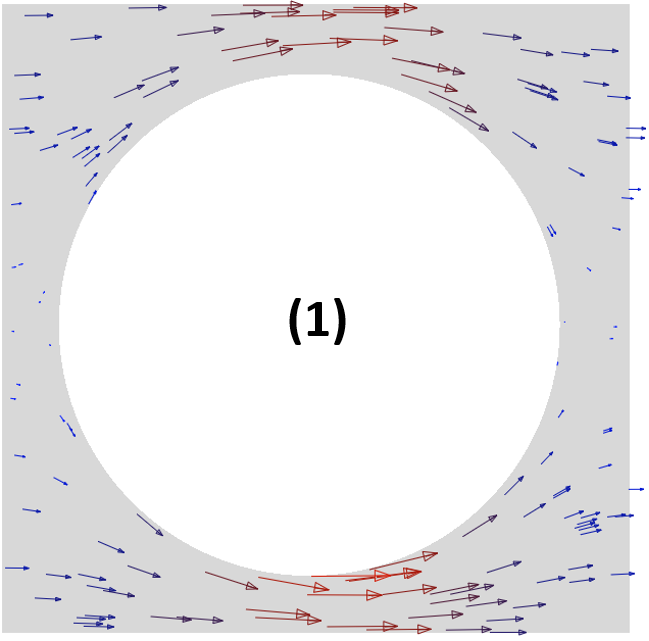} \label{circle_x}
}
\hfill
\centering
\subfigure[]{
\includegraphics[width=4cm]{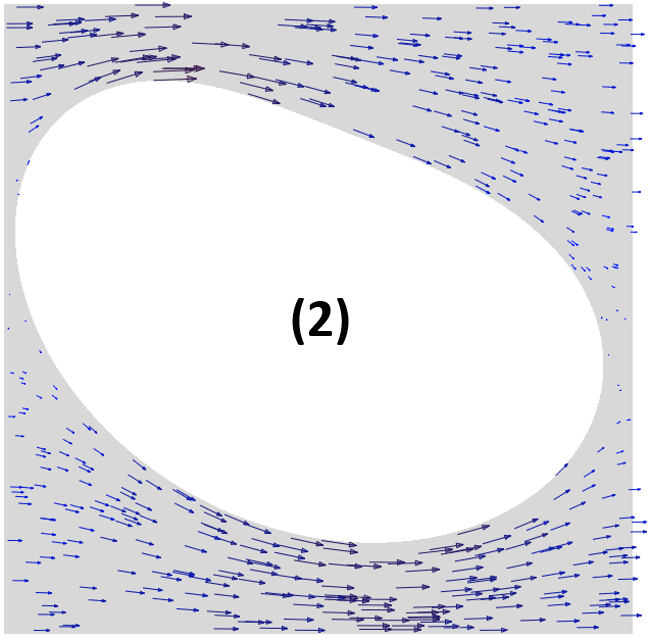}\label{a_1_b_3_x}
}
\hfill
\centering
\subfigure[]{
\includegraphics[width=4cm]{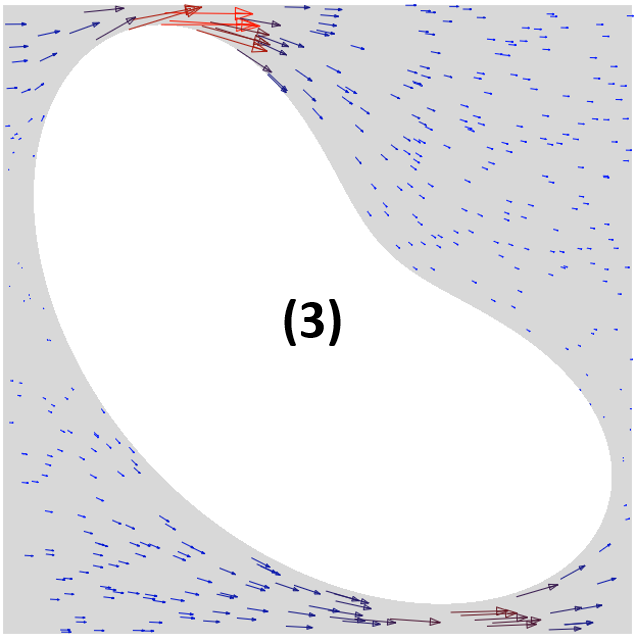}\label{a_3_b_3_x}
}
\hfill
\centering
\subfigure{
\includegraphics[width=2cm]{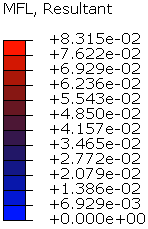}
}
\caption{Moisture mass flow rate along x direction in bean curved fiber reinforced matrix composites: (a) Case 1: a=1, b=1 or a=2, b=2 (b) Case 2: a=1, b=3 (c) Case 3: a=3, b=3}\label{MFL_x}
\end{figure}

\begin{figure}[h!]
    \centering
\subfigure[]{
\includegraphics[width=4cm]{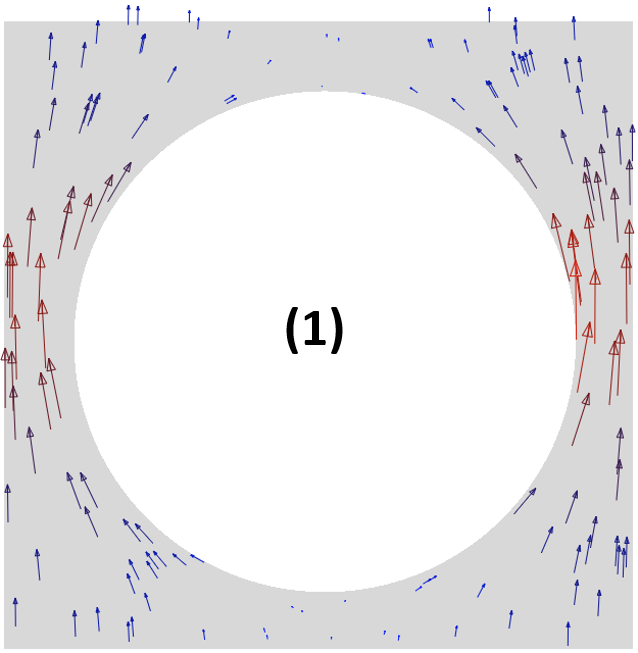} \label{circle_y}
}
\hfill
\centering
\subfigure[]{
\includegraphics[width=4cm]{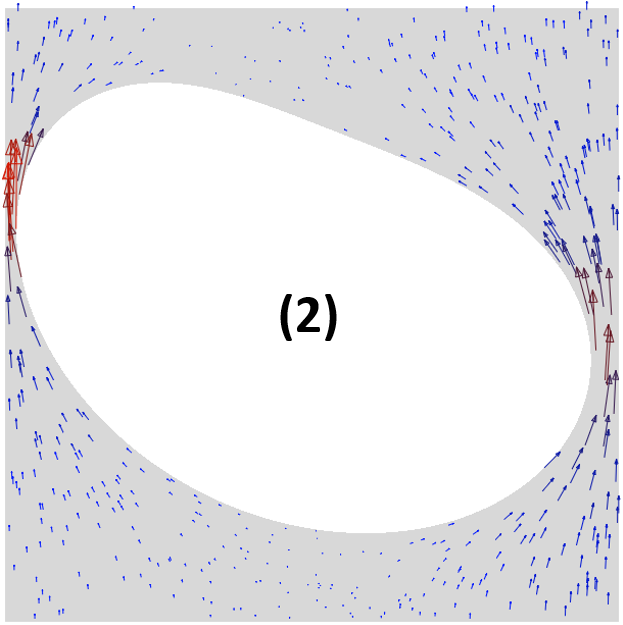}\label{a_1_b_3_y}
}
\hfill
\centering
\subfigure[]{
\includegraphics[width=4cm]{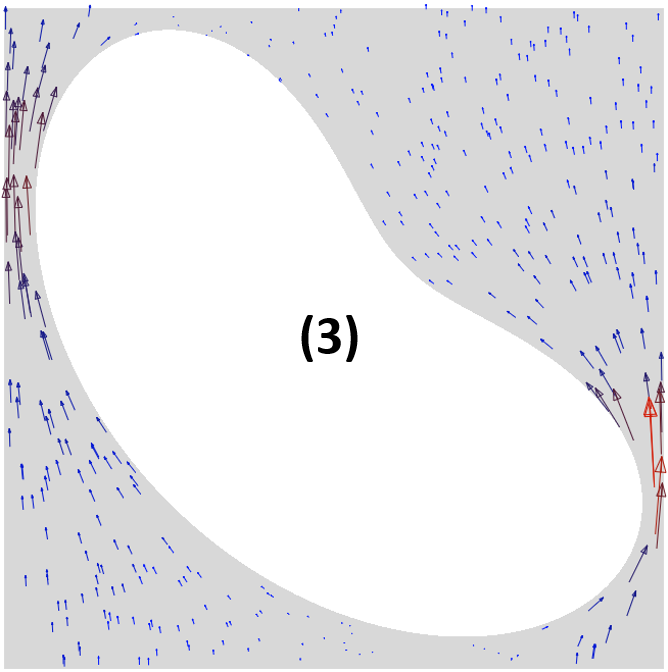}\label{a_3_b_3_y}
}
\hfill
\centering
\subfigure{
\includegraphics[width=2cm]{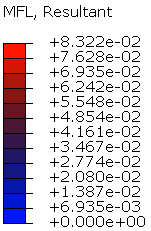}
}
\caption{Moisture mass flow rate along y direction in bean curved fiber reinforced matrix composites (a) Case 1: a=1, b=1 or a=2, b=2 (b) Case 2:a=1, b=3 (c) Case 3:a=3, b=3}\label{MFL_y}
\end{figure}

We have highlighted three cases in Figures~\ref{shape_x} and \ref{shape_y}, which correspond to three fiber cross-sectional shapes shown in Figures~\ref{MFL_x} and \ref{MFL_y}, respectively.
In Figure~\ref{shape_x}, we observe that case (2), a bean curve with a=1 and b=3, manifests the lowest tortuosity compared to all other bean curves considered. Case(1) with a=1 and b=1 (also, a=2 and b=2) is essentially a circle which manifests a tortuosity value marginally greater than case (2), but lower than the other types of bean curves considered in this study. Case (3) is a bean curve with a=3 and b=3, which shows the highest tortuosity along the x direction compared to all other fiber shapes. We plotted moisture concentration flux vector distributions for cases (1)-(3) in Figure~\ref{MFL_x} for moisture flow occurring along the x direction. We observe that the average mass flux in Figure~\ref{a_3_b_3_x} is less compared to that of Figure~\ref{circle_x} and Figure~\ref{a_1_b_3_x}, which in turn increases the tortuosity. 

For moisture flow occurring along the y direction, the circular fiber (case (1)) has lower tortuosity compared to the other two bean curves (case (2) and case (3)), as shown in Figure~\ref{shape_y}. The reason for this behavior is that the distance over which the moisture flow occurs (impact length) is higher for case (2) as can be seen in Figure~\ref{a_1_b_3_y} than that in Figure~\ref{circle_y}, which hinders the moisture ingression. However, when the moisture flow occurs along x direction for Case(2) as shown in Figure~\ref{a_1_b_3_x}, the impact length is lower to that of a circular fiber. This is further corroborated by the tortuosity values for flow along that direction. 
The real microstructure within a fiber reinforced composite might introduce higher hindrance against moisture diffusion for lesser fiber volume fraction due to the morphology, which will be discussed next.

\subsection{Influence of Real Microstructure on Tortuosity}\label{sse:micro}
\begin{figure}[H]
\centering
\subfigure[]{
\includegraphics[width=4.5cm]{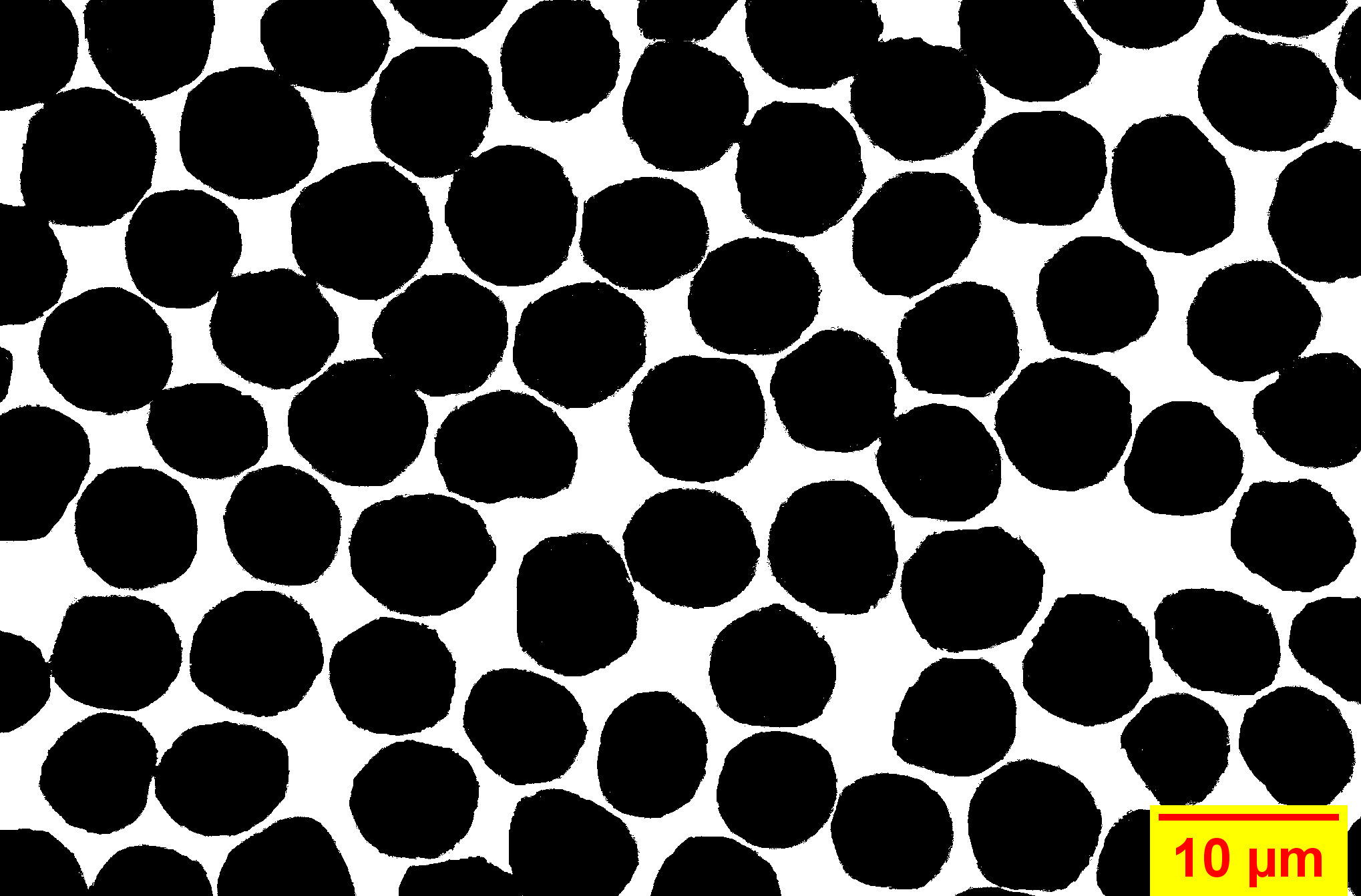} \label{real_micro_1}
}
\centering
\subfigure[]{
\includegraphics[width=4.5cm]{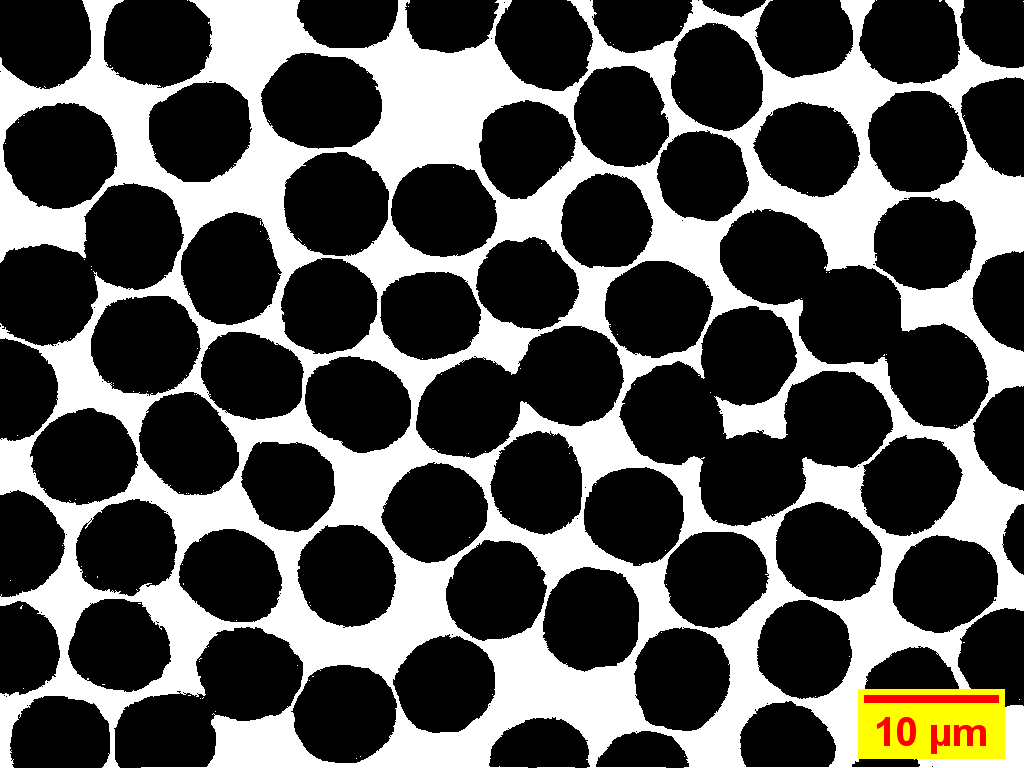}\label{real_micro_2}
}
\centering
\subfigure[]{
\includegraphics[width=4.5cm]{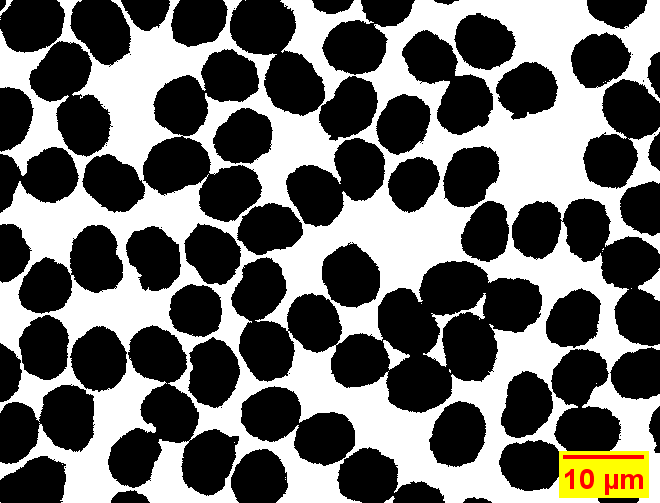}\label{real_micro_3}
}
\centering
\subfigure[]{
\includegraphics[width=3.5cm]{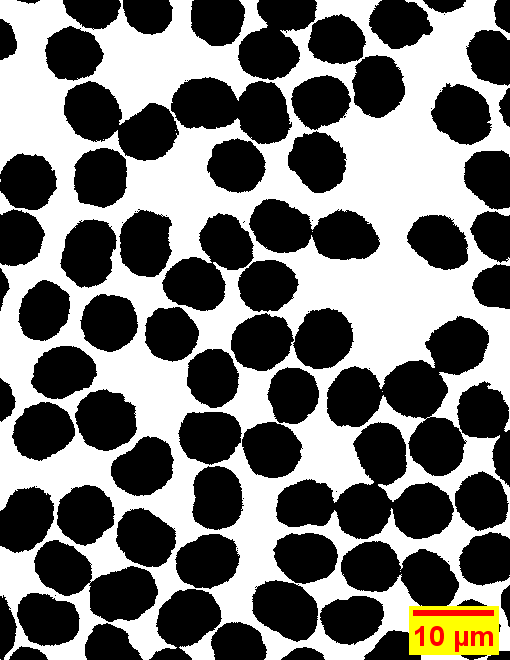}\label{real_micro_4}
}
\centering
\subfigure[]{
\includegraphics[width=4.5cm]{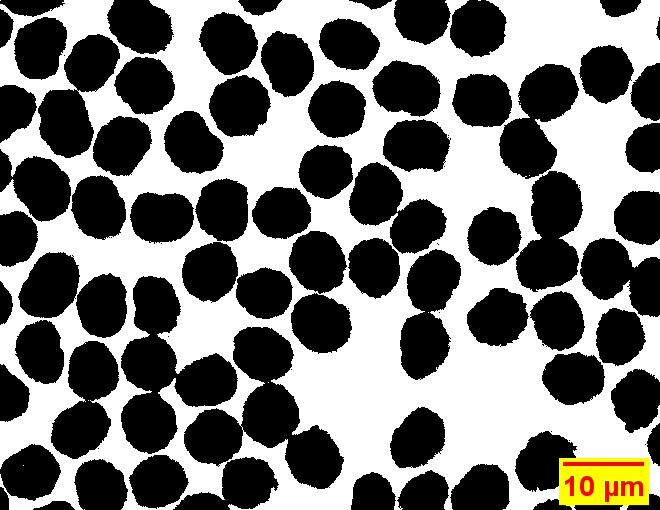}\label{real_micro_5}
}
\caption{Microstructural binary image of Carbon Fiber reinforced polymer composites}\label{real_micro_all} 
\end{figure}

\begin{figure}[h!]
\centering
\subfigure[]{
\includegraphics[width=7cm]{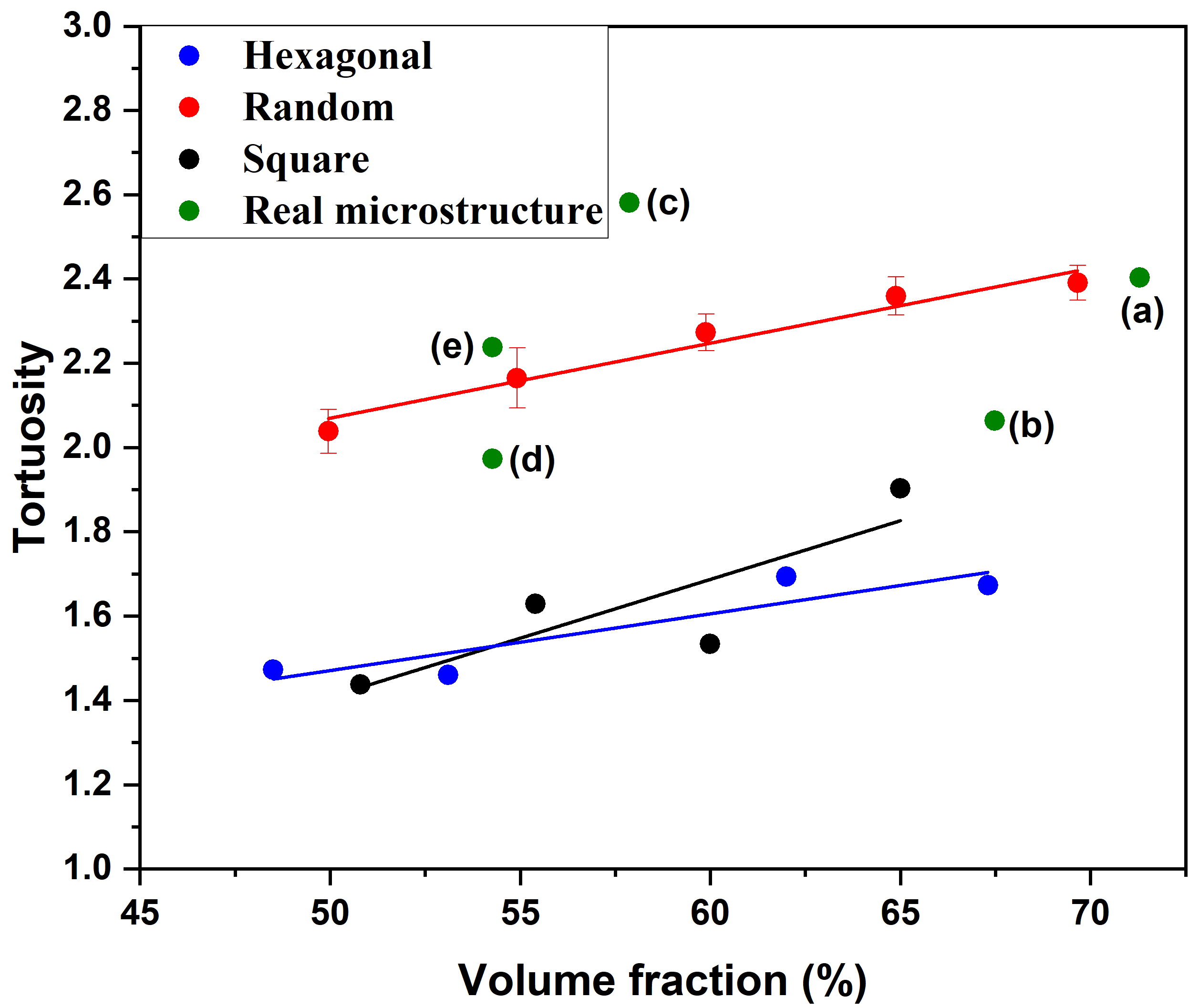} \label{tort_micro}
}
\centering
\subfigure[]{
\includegraphics[width=7cm]{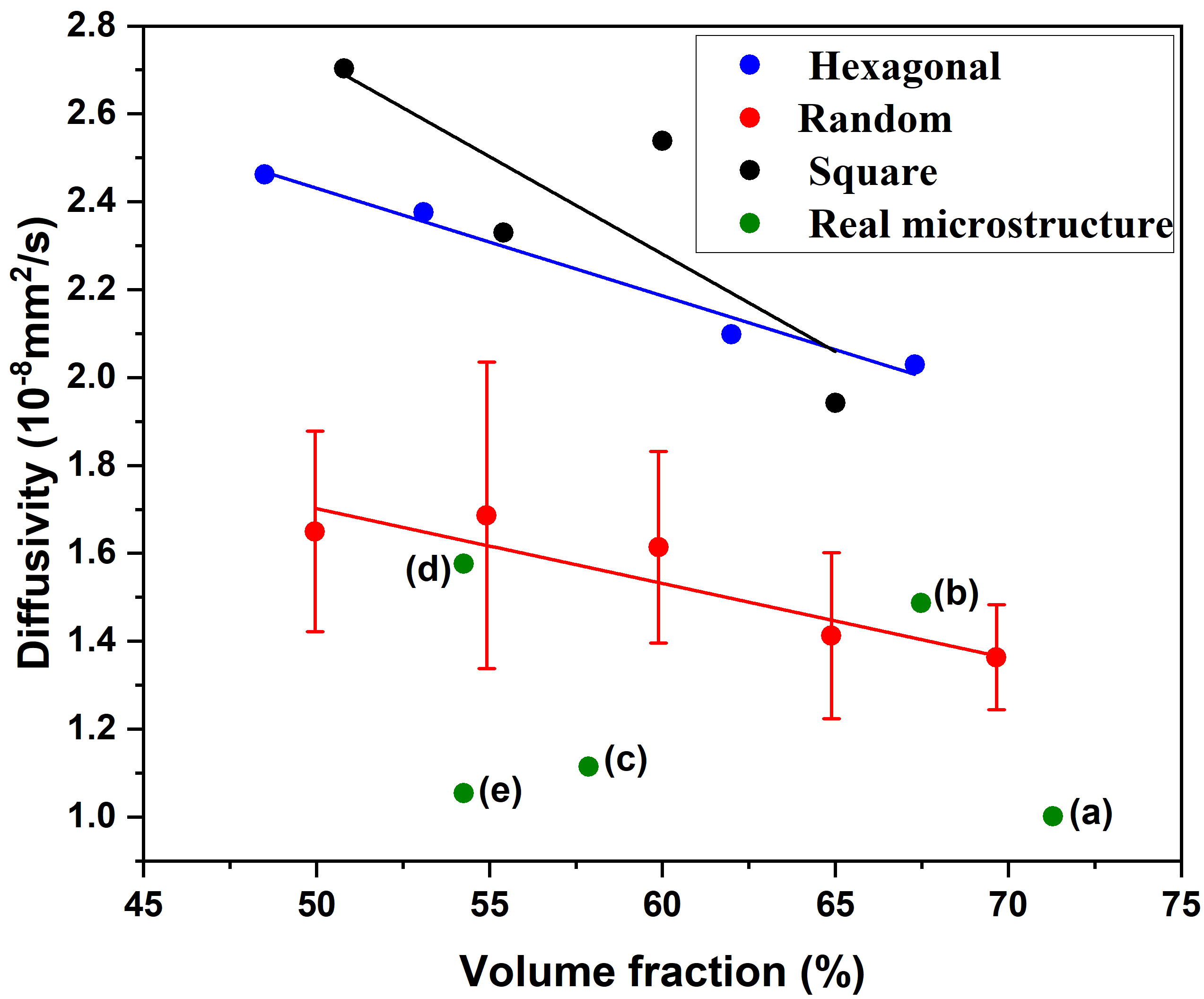}\label{diff_micro}
}
\caption{(a) Tortuosity (real microstructure) vs fiber volume fraction; (b) Diffusivity (real microstructure) vs fiber volume fraction}
\end{figure}

In sections~\ref{sse:effect_of_vf} and ~\ref{sse:effect_of_fiber_shape}, we have shown that fiber spatial distribution and morphology independently influence the composite's moisture diffusivity. Most prior work \cite{Whitcomb2002,Joliff2012a} assumed fibers to be circular and arranged in regular arrays (like square packed and hexagonal packed unit cell models) for calculating effective properties. However, real microstructure is a combination of fiber shapes and distributions in a homogeneous media. Hence, it is important to consider real microstructure for analyzing the kinetics of moisture flow in fiber reinforced composites. Here, we imported real microstructure into a finite element model as explained previously in Section~\ref{se:compModel}. Figure~\ref{real_micro_all} shows five binary images of real microstructure. Their tortuosity and diffusivity values are shown in Figures~\ref{tort_micro} and \ref{diff_micro}, respectively, along with those for computer generated random, square and hexagonal fiber arrangements.

In Figure~\ref{tort_micro}, we observe that the tortuosity values for real microstructures lie in the range of 2D micromechanical models with circular fibers. In addition, the tortuosity values for microstructures in Figures~\ref{real_micro_1} and \ref{real_micro_2} are in the same range as that for Figure~\ref{real_micro_3}, \ref{real_micro_4} and \ref{real_micro_5}, despite having higher fiber volume fractions. This behavior is attributed to fiber morphology. We observe that the concentration of non-circular (bean curve shaped) fiber cross-sections is higher in Figures~\ref{real_micro_3}, \ref{real_micro_4} and \ref{real_micro_5} than that in Figures~\ref{real_micro_1} and \ref{real_micro_2}. These fiber shapes create hindrance against moisture diffusion by increasing the tortuosity, as was discussed in section~\ref{sse:effect_of_fiber_shape}. We also observe that the tortuosity values in Figure~\ref{tort_micro} are inversely proportional to diffusivity values in Figure~\ref{diff_micro}, where the diffusivity values are higher for lower tortuosity values, and vice versa. In summary, we have elucidated the influence of fiber morphology and distribution in real microstructures on tortuosity and diffusivity in this paper.

\subsection{Relationship between Tortuosity and Diffusivity}\label{sse:relationship}

Next, we establish a relationship between tortuosity and normalized diffusivity such that we can predict diffusivity of a given microstructure by merely determining their tortuosity. Recall that, we only need to perform a steady state diffusion analysis to determine tortuosity. To that end, we analyzed more than hundred two-dimensional micromechanical models with different fiber architecture that includes fiber distribution, volume fraction and fiber morphology, and determined their diffusivity and tortuosity values. The tortuosity and diffusivity are calculated using the method mentioned in Section~\ref{diffusionModeling}. We define a "normalized diffusivity" term, which is the ratio of effective diffusivity ($D_{eff}$) to that of homogeneous matrix diffusivity ($D_m$). Then, we performed a regression analysis to establish a relationship between normalized diffusivity and tortuosity. This relationship is shown in Equation~\mbox{\ref{eq:relationship}}:
\begin{equation}\label{eq:relationship}
\begin{split}
    y = y_0 + A* exp(R_0*x)\\
    y_0 = 0.08503 \pm 0.00564 \\
    A =   1.95986 \pm 0.90726\\
    R_0 = -2.3928 \pm 0.39118 
    \end{split}
\end{equation}

where, $y$ is the normalized diffusivity and x is the tortuosity which is dimensionless. The fitting constants $y_0$, $A$ and $R_0$ are the offset, initial value and rate, respectively, and are also dimensionless. Here, $R_0$ denotes the rate of reduction in the normalized diffusivity with increasing tortuosity as it has a negative value. As the fiber volume fraction decreases to 20\% and 30\%, the difference in diffusivity values is higher than that of tortuosity values. This can be attributed to an increased variability in the fiber packing at lower volume fractions. Whereas, this variability decreases for higher volume fractions where the fiber packing converges towards a uniform hexagonal packing. This is captured well with $R_0$ in this equation.
 Using a normalized diffusivity enables us to generalize this relationship to other similarly behaving FRPC materials. Since, a majority of composite materials for structural applications consist of fiber volume fractions ranging between 20\text{\%} to 70\text{\%}, models with only these volume fractions are considered in the regression analysis. 

\begin{figure}[h!]
 \centering
  \includegraphics[width=10cm]{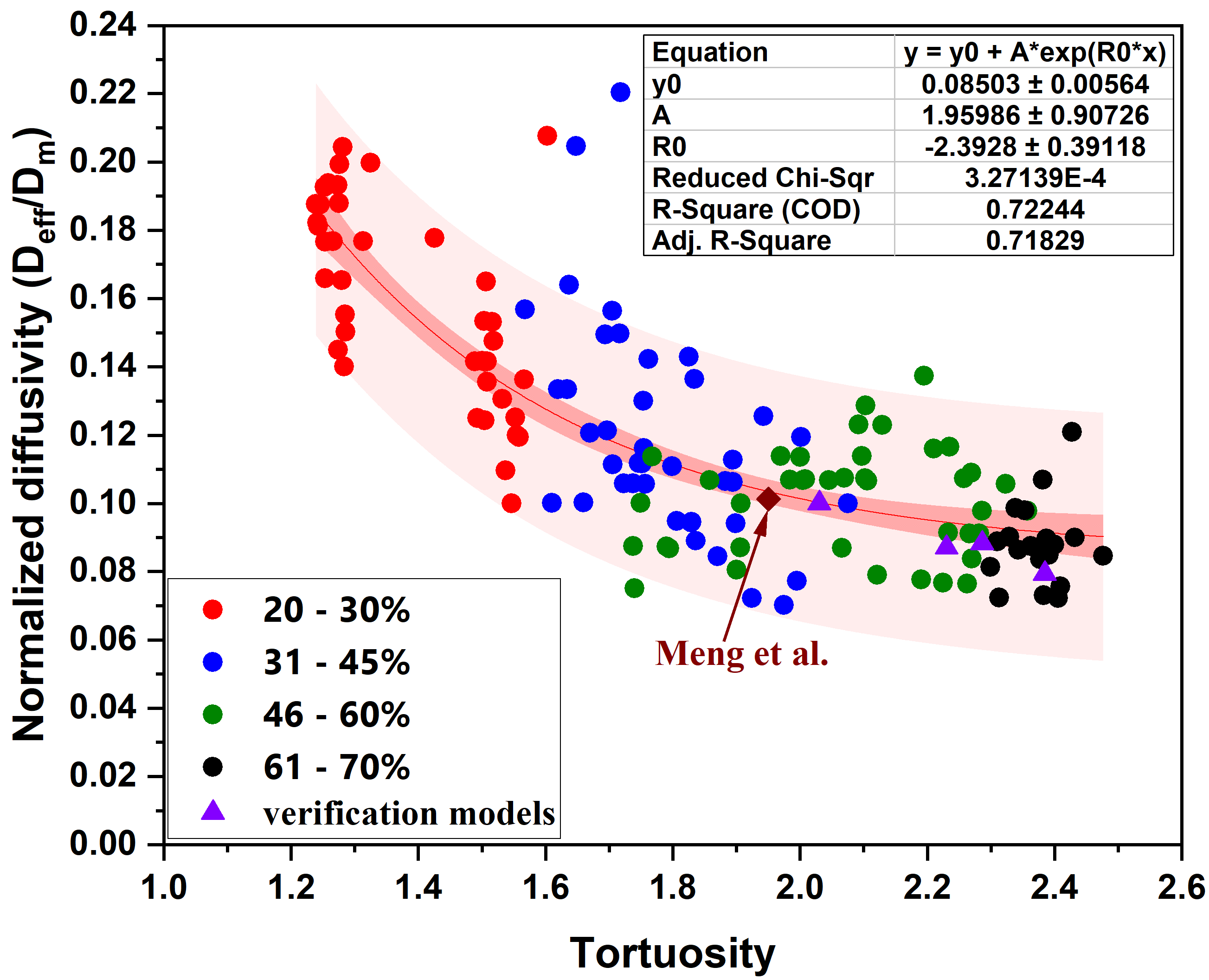}
  \caption{Relationship between tortuosity and normalized diffusivity considering a combination of circular and real microstructure models }\label{relationship_plot}
\end{figure}

Figure~\ref{relationship_plot} shows the tortuosity and corresponding normalized diffusivity ($D_{eff}/D_m$) for several chosen models. The relationship between tortuosity and normalized diffusivity is an exponential fit with an $R^2$ value of 0.72. We show again from Figure~\ref{relationship_plot} that diffusivity decreases with increasing tortuosity, which in turn increases the time required for moisture to reach its saturation limit in a given fiber-reinforced composite. We have also highlighted the diffusivity/tortuosity values using different colors (blue, green and red) for certain ranges of fiber volume fractions. As expected, diffusivity decreases and tortuosity increases in general as fiber volume fraction increases. However, other factors like fiber distribution and morphology introduce overlapping transition zones as seen in Figure~\ref{relationship_plot}.

In Figure~\ref{relationship_plot}, we see that there is some scatter in the data points around the fit, which we attribute to the well established Fick's law used for calculating the moisture diffusivity. That is, the weight gain is considered linear in the initial portion, which is followed by a non-linear region prior reaching saturation. However, we observe a small deviation from this behavior in the initial portion of our computational results, that is, there is a sudden rise initially and then a more gradual linear weight gain portion. This is attributed to Fick's law which assumes that when a sample comes in contact with moisture, equilibrium is instantaneously established between environmental moisture and internal moisture concentration. However, this process is gradual in reality \cite{weitsman,Pilli2014}. This dual initial slope from the simulations can introduce some variation in how the initial slope of the weight gain curve is calculated, which can manifest as fluctuations in the calculated diffusivity values.

Finally, for validating the proposed tortuosity-normalized diffusivity relationship, we use four different randomly distributed fiber architectures along with a fiber architecture model presented in Meng et al. \cite{Meng2016}. We determined the tortuosity values using finite element method. Then, with the tortuosity values as input, we calculated the normalized diffusivity values using the established relationship. Our validation points fall within a 95\% confidence band of the established relationship. 

\section{Conclusions}

Key outcomes of this study are:
\begin{enumerate}
    \item We elucidated the influence of fiber architecture, which in 2D space includes fiber volume fraction, spatial fiber distribution and fiber morphology, on the effective moisture diffusivity of fiber reinforced composites by presenting a unifying concept of tortuosity. The tortuosity factor captured the effective fiber architecture.
	\item We developed micromechanical models consisting of random,square and hexagonal array with varying fiber volume fraction (50-70\text{\%}) to study the influence of volume fraction and distribution on tortuosity and diffusivity. We noted that an increase in fiber volume fraction increases the tortuosity, thereby, decreasing its moisture diffusivity. We also showed that the fiber distribution has an impact on tortuosity and moisture diffusivity, where the models with randomly distributed fibers manifested higher tortuosity and lower diffusivity compared to that of the regular array models.
	\item We investigated the impact of fiber morphology on tortuosity using a unit cell model with fixed (50\text{\%}) fiber volume fraction, where perfectly circular fibers mostly underestimated the tortuosity values and overestimated the diffusivity values.
	\item To understand the combined effect of fiber architecture on tortuosity and diffusivity, we considered real microstructures of CFRP composites. Here, we showed that models with lower fiber volume fractions can manifest higher tortuosity due to the combined influence of fiber morphology and distribution as compared to randomly distributed models with circular fibers.
	\item We showed that tortuosity and moisture diffusivity are inversely proportional in fiber reinforced composites. We established a relationship between tortuosity and diffusivity from which diffusivity can be estimated using tortuosity. It is worth noting that tortuosity can be easily calculated by solving steady state diffusion governing equation as compared to time dependent transient diffusion governing equation required to solve for moisture diffusivity. Hence, this can be very valuable for designing fiber architectures, multi-scale analysis and optimization. 
	\item In summary, we elucidated the influence of fiber architecture on tortuosity, which is inversely related to moisture diffusivity. Using this knowledge, we can devise fiber architectures with greater tortuosity such that it results in increased moisture saturation time, which in turn is expected to increase the time required for moisture driven degradation. Thus, enabling longevity of fiber reinforced composites.
\end{enumerate}

\section*{Acknowledgements}
The authors would like to acknowledge the support through the U.S. Office of Naval Research - Young Investigator Program (ONR-YIP) award [N00014-19-1-2206] through {\em{Sea-based Aviation: Structures and Materials Program}} for conducting the research presented here. 

\section*{Data Availability}

The raw/processed data required to reproduce these findings cannot be shared at this time as the data also forms part of an ongoing study.



\bibliographystyle{unsrt}
\bibliography{Ref}

\end{document}